\documentclass[11pt,a4paper]{article}
\usepackage{jstyle}

\usepackage{tikz}
\usepackage{verbatim}
\usetikzlibrary{fit}
\tikzset{%
  highlight/.style={rectangle,rounded corners,fill=red!15,draw,fill opacity=0.5,thick,inner sep=0pt}
}
\newcommand{\tikzmark}[2]{\tikz[overlay,remember picture,baseline=(#1.base)] \node (#1) {#2};}
\newcommand{\Highlight}[1][submatrix]{%
    \tikz[overlay,remember picture]{
    \node[highlight,fit=(left.north west) (right.south east)] (#1) {};}
}

\def\a{\alpha}
\def\b{\beta}
\def\g{\gamma}

\def\d{\delta}

\def\L{\Lambda}
\def\m{\mu}
\def\n{\nu}

\def\r{\rho}

\def\s{\sigma}

\def\t{\tau}

\def\pr{\partial}

\def\cA{{\cal A}}

\def\cC{{\cal C}}
\def\cD{{\cal D}}

\def\cF{{\cal F}}

\def\cI{{\cal I}}

\def\cK{{\cal K}}
\def\cL{{\cal L}}

\author{Euihun JOUNG \quad}
\author{Karapet MKRTCHYAN\footnote{On leave from Yerevan Physics Institute.}}

\affiliation{Scuola Normale Superiore and INFN\\
Piazza dei Cavalieri 7, 56126 Pisa, Italy}

\emailAdd{euihun.joung@sns.it}
\emailAdd{karapet.mkrtchyan@sns.it}

\title{\centering A note on higher-derivative actions \\
for free higher-spin fields}

\abstract{Higher-derivative theories
of free higher-spin fields are investigated focusing on their symmetries.
Generalizing familiar two-derivative constrained formulations,
we first construct \emph{less-constrained} Einstein-like and Maxwell-like
higher-derivative actions.
Then, we construct Weyl-like actions---the actions admitting \emph{constrained} Weyl symmetries---with different numbers of derivatives.
They are presented in a factorized form making use of Einstein-like and
Maxwell-like tensors.
The last (highest-derivative) member of the hierarchy of the Weyl-like actions
coincides with the Fradkin-Tseytlin conformal higher-spin action in four dimensions.
}

\begin{document}

\maketitle

\section{Introduction and Summary}
\label{sec: intro}

Massless higher-spin representations of the Poincar\'e group
require higher-rank tensor gauge fields for their description.
However, from a geometrical viewpoint,
these fields have an unusual field-theoretical implication
since they favor higher derivatives, and for instance
the generalized curvature for a rank-$s$ gauge field
involves $s$ derivatives \cite{deWit:1979pe}.
As a result, the Lagrangian equation of motion constructed from the curvature would propagate not only spin-$s$ modes but also additional ones
including ghosts.
For spin-$s$ propagation,
one has to reduce the number of derivatives to two:
this can be done either by imposing constraints on gauge fields and parameters \cite{Fronsdal:1978rb}, or by introducing auxiliary fields \cite{Francia:2005bu,Buchbinder:2007ak} or non-locality \cite{Francia:2002aa}.
See e.g. \cite{Bekaert:2005vh,Bekaert:2010hw,Sagnotti:2011qp}
for recent reviews on higher-spin gauge theories.

Apart from the higher-spin context,
higher-derivative theories have been considered mainly
for the purpose of renormalizable gravity \cite{Stelle:1976gc}.
Again, the downside is the non-unitary dynamics with propagating ghost modes.
Recently higher-derivative theories started to attract renewed interest, thanks to several observations on how to deal with the ghost problem.
Among them, let us mention the following key points:
\begin{itemize}
\item
In  low dimensions, such ghosts can
become pure gauge restoring unitarity.
An important example is the New Massive Gravity
\cite{Bergshoeff:2009hq}.\footnote{
See also \cite{Bergshoeff:2009tb,Bergshoeff:2011pm,Bergshoeff:2012ud}
 and references therein for generalizations.}
\item
The solutions of Einstein gravity with negative cosmological constant
can be recovered from certain four-derivative gravity theories
choosing appropriate boundary conditions (see e.g. \cite{Metsaev:2007fq,
Maldacena:2011mk,Lu:2011ks,Hyun:2011ej}).
\end{itemize}
It is worth noticing that the Weyl symmetry underlies the consistency of these four-derivative theories,
although the Lagrangian is not Weyl invariant in general.
More precisely, in all these theories, the four-derivative part of the action
enjoys a Weyl symmetry at least at the linearized level.
In a sense, this symmetry removes
the massive scalar from the spectrum, leaving only massless and massive spin two modes -- which are relatively ghost.\footnote{General four-derivative gravity propagates
a massless spin two, a massive scalar and a massive spin two ghost \cite{Stelle:1976gc,Stelle:1977ry,Lee:1982cp,Buchbinder:1987vp}.}
It is instructive to see this point in detail.
Around a constant-curvature background
$\bar g_{\mu\nu}$\,, linearized Weyl gravity
admits the factorized expression \cite{Deser:2012qg}
\ba\label{hd gravity}
	S_{1}=\int d^{d}x\sqrt{-\bar g}\, \left[
	G^{\ \mu\nu}_{\rm\sst lin}\,(I^{\ -1}_{\sst\rm FP})_{\mu\nu,\rho\sigma}\,
	G^{\ \rho\sigma}_{\rm\sst lin}+\tfrac{d-2}{2(d-1)}\,
	\Lambda\,h^{\mu\nu}\,G^{\sst\rm lin}_{\, \mu\nu}\right],
\ea
in terms of the Fierz-Pauli (FP) mass operator
\be
(I_{\sst\rm FP})_{\mu\nu,\rho\sigma}=
    \bar g_{\mu\rho}\,\bar g_{\nu\sigma}-\bar g_{\mu\nu}\,\bar g_{\rho\sigma}
\ee
Where
$G^{\sst\rm lin}_{\, \mu\nu}=R_{\mu\nu}-\frac{1}{2}\,\bar g_{\mu\nu}\,R+\L\,\bar g_{\mu\nu}$ is the linearized cosmological Einstein tensor.
Thanks to the FP operator in $S_{1}$\,, with the addition of a massless spin two action $S_{2}$\,:
\be
	S_{2}= m^{2}\int d^{d}x\sqrt{-\bar g}\ h^{\mu\nu}\,G^{\sst\rm lin}_{\, \mu\nu}\,,
\ee
the full action can be easily recast into the \emph{difference} between
massless and massive spin two actions:
\be\label{spin 2 m}
	S_{1}+S_{2} = M^{2}\int d^{d}x\sqrt{-\bar g}\
	\Big[\,\tilde h^{\mu\nu}\,G^{\sst\rm lin}_{\, \mu\nu}(\tilde h)
	-\varphi^{\mu\nu} \Big\{ G^{\sst\rm lin}_{\, \mu\nu}(\varphi)
	+M^{2}\,(I_{\sst\rm FP})_{\mu\nu}^{\rho\sigma}\,\varphi_{\rho\sigma}
	\Big\}\Big],
\ee
with \mt{M^{2}=m^{2}+\tfrac{d-2}{2(d-1)}\,\Lambda}\,.
The equality in \eqref{spin 2 m} can be proved by first performing field redefinition \mt{\tilde h_{\mu\nu} = h_{\mu\nu}-\varphi_{\mu\nu}}\, and then solving $\varphi_{\mu\nu}$ in terms of \mt{h_{\mu\nu}}.
In three dimensions, the massless spin two does not propagate, and
therefore with an appropriate choice of the sign for the kinetic term
one can retain only ghost-free massive spin two.
On the other hand, in generic dimensional AdS backgrounds,
the massless spin two can be selected as the only propagating mode
imposing appropriate boundary conditions.

Turning to the case of higher-spin fields,
one may wonder whether Weyl(-like) symmetries can still play
a similar  role in controlling the spectrum of
four(or even higher)-derivative theories.
 Focusing on the free case,
the natural generalization of (linearized) diffeomorphisms and Weyl transformations
to higher spins would be
\be\label{gw tr}
	\delta\,\varphi_{\mu_{1}\cdots \mu_{s}}=
	\partial_{(\mu_{1}}\,\varepsilon_{\mu_{2}\cdots\mu_{s})}
	+\eta_{(\mu_{1}\mu_{2}}\,\alpha_{\mu_{3}\cdots\mu_{s})}\,,
\ee
where $(\mu_{1}\cdots \mu_{s})$ denotes full symmetrization.
In $d\geq 4$\,, a free action possessing the above symmetry
involves the square of the higher-spin Weyl tensor
(the traceless part of the higher-spin curvature),
and thus contains $2s$ derivatives for a given spin $s$\,.
For \mt{d=4}\,, it coincides with the conformal higher-spin theory\footnote{
Conformal spin-$s$ actions in $d$ dimensions contain \mt{(d+2s-4)}-derivatives.
In order to avoid confusion between the conformal actions and the $2s$-derivative Weyl squared actions,
we call the latter Weyl action, as opposed to conformal action,
 throughout the present paper.} of
 \cite{Fradkin:1985am,Metsaev:2007rw,Marnelius:2008er} (see also \cite{Shaynkman:2004vu,Metsaev:2008fs,Metsaev:2009ym,Vasiliev:2009ck,Bekaert:2012vt}).
Although one can directly consider $2s$-derivative theories
and study their ghost exorcising mechanisms, one can still wonder whether
interesting theories exist whose number of derivatives
lies between 2 and $2s$\,.

\medskip

\begin{table}[hbp]
\label{table}
\centering
  \begin{tabular}{ p{1.5cm} | p{1.5cm} | p{1.5cm} | p{1.5cm} | p{1.5cm} }
    \centering $\partial^{2}$ & \centering $\partial^{4}$ &
    \centering $\cdots$ & \centering $\partial^{2s-2}$ &  \quad\  $\partial^{2s}$
    \\ \hline
    Fronsdal & \centering? & \centering$\cdots$ & \centering? &\ \ \,Weyl
  \end{tabular}
  \caption{Hierarchy of higher-derivative actions for higher spins}
\end{table}
\noindent
If these theories exist, we expect that the more derivatives
they contain, the more their symmetries are enhanced. The two ends of this hierarchy of theories correspond to the Fronsdal
and higher-spin Weyl theories that admit, respectively,
constrained gauge symmetries and unconstrained gauge plus Weyl symmetries, while other members are expected to possess symmetries
that are intermediate between these.

\bigskip

In the present paper we investigate higher-derivative theories
of free higher-spin fields focusing on mainly their symmetries.
We begin with familiar two-derivative constrained formulations of higher-spin fields: Fronsdal's theory \cite{Fronsdal:1978rb} and transverse invariant setting of Skvortsov and Vasiliev (SV) \cite{Skvortsov:2007kz}, recently investigated in a wider context in \cite{Campoleoni:2012th}.
Then, increasing the number of derivatives, we identify the Lagrangians
compatible with relaxed constraints, and in particular those
acquiring Weyl(-like) symmetries.

\paragraph{Einstein-like actions}

We find that increasing the number of derivatives
from Fronsdal's setting (doubly traceless field: \mt{\varphi''=0}\,, and traceless parameter: \mt{\varepsilon'=0}),
at each step there is unique Lagrangian with higher trace constraints.
More precisely, the $2n$-derivative Einstein-like actions are determined as
\be
	\mathscr G_{\sst 2n}=\int d^{d}x\ \varphi_{\mu_{1}\cdots \mu_{s}}
	G^{\mu_{1}\cdots \mu_{s}}_{\sst 2n}\,,
\ee
where the gauge field is subject to the \mt{(n+1)}-th
traceless constraint: \mt{\varphi^{[n+1]}=0}\,,
while the gauge parameter to the $n$-th
traceless constraint: \mt{\varepsilon^{[n]}=0}\,.

\paragraph{Maxwell-like actions}
On the other hand, departing from the setting of SV
(traceless field: \mt{\varphi'=0}\, and traceless and transverse parameter:
\mt{\varepsilon'=0=\partial\cdot\varepsilon}), we obtain the
$2n$-derivative Maxwell-like actions
\be
	\mathscr M_{\sst 2n}=\int d^{d}x\ \varphi_{\mu_{1}\cdots \mu_{s}}
	M^{\mu_{1}\cdots \mu_{s}}_{\sst 2n}\,,
\ee
whose fields and parameters are subject to the constraints:
\mt{\varphi^{[n]}=0} and \mt{\varepsilon^{[n]}=0=\partial\cdot\varepsilon^{[n-1]}}.
Let us notice that the Maxwell-like actions $\mathscr M_{\sst 2n}$
can be regarded as partially gauge fixed versions of the
Einstein-like actions $\mathscr G_{\sst 2n}$\,,
as its two-derivative case (\mt{n=1}).

It is worth noticing that the Einstein-like
tensors $G_{\sst 2n}$ and  the Maxwell-like tensors $M_{\sst 2n}$ admit
a full factorization in terms of $n$ two-derivative operators,
whose first factors are the Fronsdal and Maxwell ones.
It was shown in \cite{Francia:2012rg}
that the equations \mt{G_{\sst 2n}\approx 0} and \mt{M_{\sst 2n}\approx 0}
are equivalent respectively to the equations \mt{R^{\sst [n]}\approx 0} and \mt{\partial\cdot R^{\sst [n-1]}\approx 0}
where $R$ is the linearized higher-spin curvature.
These results accord with the conjecture
based on the cohomological analysis of the equations of motion (see the footnote 25 in \cite{Bekaert:2005vh}).

The Einstein-like actions exist up to \mt{2\lfloor s/2\rfloor} derivatives
while the Maxwell-like ones to \mt{2\lfloor (s+1)/2 \rfloor} derivatives.
Hence, they can cover only half of Table \ref{table}.
The last member of each set (Einstein-like for even spin and Maxwell-like for odd spin) admits an unconstrained gauge symmetry and
is essentially the local counterpart of
the unconstrained non-local Lagrangian of \cite{Francia:2002aa}.
Although there is no generic symmetry enhancement beyond
$s$ derivatives, we show that for \mt{2s+d-4} derivatives
Einstein-(Maxwell-)like actions acquire Weyl symmetry.
They coincide with the linearized conformal higher-spin action
since the latter has the same number of derivatives and the same symmetries.

\paragraph{Weyl-like actions}

The higher-spin Weyl action
involves $2s$ derivatives and does not require any constraint on its gauge field and parameters.
It turns out that Weyl symmetry can be realized
also in lower derivative  cases (from four to \mt{2s-2} derivatives) with constrained parameters.
We identify Lagrangians possessing Weyl symmetries with parameters, subject to differential constraints.
The Weyl-like actions are split into two classes:
the first involves $4n$ derivatives
while the other involves \mt{4n+2} derivatives.
\begin{itemize}
\item
The $4n$-derivative Weyl-like actions are obtained through Einstein-like tensors as
\be\label{W4n}
	\mathscr W_{\sst 4n}=\int d^{d}x\ G^{\mu_{1}\cdots \mu_{s}}_{\sst 2n}\,
	(I^{\,-1}_{\sst\rm F})_{\mu_{1}\cdots\mu_{s},\nu_{1}\cdots\nu_{s}}\,
	G^{\nu_{1}\cdots\nu_{s}}_{\sst 2n}\,,
\ee
where $I_{\sst\rm F}$ is the mass operator for higher-spin fields
considered by Francia \cite{Francia:2007ee}.
Let us notice the similarity between
these Lagrangians \eqref{W4n} and the gravity case \eqref{hd gravity}.
Both cases admit a factorization in terms of two Einstein(-like) tensors
with the inverse of the Fierz-Pauli (Francia) mass operator.
The constraints on gauge fields and parameters follow those
of the $2n$-derivative Einstein-like action $\mathscr G_{\sst 2n}$\,,
while the Weyl parameters are subject to new types of constraints --
traceless and $(2n-1)$-th divergence-less conditions:
\mt{\alpha'=0=(\partial\cdot)^{2n-1}\alpha}\,.
\item
The \mt{(4n+2)}-derivative Weyl-like actions are determined using
 both Einstein-like and Maxwell-like tensors, as
\be
	\mathscr W_{\sst 4n+2}=\int d^{d}x\ G^{\mu_{1}\cdots \mu_{s}}_{\sst 2n}\,
	(I_{\sst\rm F}^{\,-1})_{\mu_{1}\cdots\mu_{s},\nu_{1}\cdots\nu_{s}}\,
	M^{\nu_{1}\cdots\nu_{s}}_{\sst 2n+2}\,.
\ee
In this case, the constraints on the gauge fields and parameters follow those
of the \mt{(2n+2)}-derivative Maxwell-like action $\mathscr M_{\sst 2n+2}$\,,
while the Weyl parameters satisfy \mt{\alpha'=0=(\partial\cdot)^{2n}\alpha}\,.
\end{itemize}
Note that these Weyl-like actions coincide with the higher-spin Weyl action
when \mt{n=s/2} or $(s-1)/2$,
providing a new expression for the Weyl squared action.

\bigskip

The organization of this paper is as follows. In Section \ref{sec: gauge inv},
we review the generalized Christoffel symbols \`a la de Wit and Freedman and
construct higher-derivative Einstein-like and Maxwell-like actions for higher-spin fields. In Section \ref{sec: Weyl}, we turn to Weyl-like actions
and show that they are given in terms of Einstein-like and Maxwell-like tensors with a factor that is an inverse mass operator.
Section \ref{sec: discussion} contains the discussions
on the (A)dS deformations of the Weyl(-like) actions
and their expected properties,
as well as some issues on the interacting cases.
Finally, Appendices \ref{sec: s3 Weyl} and \ref{sec: spec} provide the explicit form of the spin three Weyl action
and the spectrum analysis of the four-derivative Einstein (or Maxwell)-like action.

\section{Higher-derivative gauge invariant actions}
\label{sec: gauge inv}

In this section we discuss higher-derivative actions which admit
only gauge symmetries (gauge fields transforming with the gradient of the parameter).
They are constructed using the deWit-Freedman generalized
Christoffel symbols (GCS)
\cite{deWit:1979pe}, which we review briefly in the following subsection.
Before starting our discussion,
let us introduce the generating function (or auxiliary variable)
notation for the higher-spin fields
\be\label{gen fn}
	\varphi^{\sst (s)}(x,u)=\tfrac1{s!}\,\varphi_{\mu_{1}\cdots\mu_{s}}(x)\,
	u^{\mu_{1}}\,\cdots\,u^{\mu_{s}}\,,
\ee
which we shall use throughout the present paper.
In this notation, the gauge and Weyl transformations \eqref{gw tr} are expressed as
\be\label{gw transf}
	\delta\,\varphi^{\sst (s)}(x,u)=
	u\cdot\partial_{x}\,\varepsilon^{\sst (s-1)}(x,u)+u^{2}\,
	\alpha^{\sst (s-2)}(x,u)\,,
\ee
where $\varepsilon^{\sst (s-1)}(x,u)$ and $\alpha^{\sst (s-2)}(x,u)$ are the generating functions
of the gauge parameters $\varepsilon_{\mu_{1}\cdots \mu_{s-1}}$
and the Weyl parameters $\alpha_{\mu_{1}\cdots \mu_{s-2}}$ defined
analogously to \eqref{gen fn}.
Moreover, the actions of free higher-spin fields
will be represented using the scalar product
\ba
	\ppd{\phi^{\sst (s)}}{\psi^{\sst (s)}} \edf
	\int d^{d}x\ e^{\partial_{u_{1}}\!\cdot\partial_{u_{2}}}\,
	\phi^{\sst (s)}(x,u_{1})\,\psi^{\sst (s)}(x,u_{2})\,\big|_{u_{1}=u_{2}=0}
	\nn
	\eq \int d^{d}x\ \tfrac1{s!}\,\
	\phi_{\mu_{1}\cdots\mu_{s}}(x)\,
	\psi^{\mu_{1}\cdots\mu_{s}}(x)\,.
\ea

\subsection{deWit-Freedman generalized Christoffel symbols}

The Christoffel symbol of gravity, linearized
around the flat background so that \mt{g_{\mu\nu}=\eta_{\mu\nu}-h_{\mu\nu}}\,,
is given by
\be
	\Gamma_{\mu\nu,\rho}=\tfrac12\,
	(\partial_{\rho}\,h_{\mu\nu}-\partial_{\mu}\,h_{\nu\rho}-\partial_{\nu}\,h_{\mu\rho})\,.
\ee
Rewriting this in the auxiliary-variable notation gives
\be\label{1st gamma}
	\Gamma^{\sst (s,1)}(x,u,v)=
	\Gamma_{\mu\nu,\rho}(x)\,u^{\mu}\,u^{\nu}\,v^{\rho}=
	(v\cdot\partial_{x}-u\cdot\partial_{x}\,
	v\cdot\partial_{u})\,
	\varphi^{\sst (s)}(x,u)\,,
\ee
with \mt{s=2}\,. The general spin case of \eqref{1st gamma}
corresponds to the first member of the GCS hierarchy,
and its variation under the gauge transformation
results in the double gradient as
\be
	\delta\,\Gamma^{\sst (s,1)}(x,u,v)
	=-\,(u\cdot\partial_{x})^{2}\,(v\cdot\partial_{u})\,
	\varepsilon^{\sst (s-1)}(x,u)\,,
\ee
while the other members of the hierarchy can be determined recursively by
\be
	\Gamma^{\sst (s,r)}(x,u,v)
	=(v\cdot\partial_{x}-\tfrac1r\,u\cdot\partial_{x}\,
	v\cdot\partial_{u})\,
	\Gamma^{\sst (s,r-1)}(x,u,v)\,,
\ee
in such a way that their gauge variations give rise to the multiple gradients
\be
	\delta\,\Gamma^{\sst (s,r)}(x,u,v)
	=\tfrac{(-1)^{r}}{r!}\,(u\cdot\partial_{x})^{r+1}\,(v\cdot\partial_{u})^{r}\,
	\varepsilon^{\sst (s-1)}(x,u)\,.
	\label{Gamma tr}
\ee
In this way, the last member of the hierarchy, called deWit-Freedman curvature,
\be
	\Gamma^{\sst (s,s)}(x,u,v)=
	\tfrac1{s!}\,(u\cdot\partial_{x}\,v\cdot\partial_{w}-
	v\cdot\partial_{x}\,u\cdot\partial_{w})^{s}\,\varphi^{\sst (s)}(x,w)\,,
\ee
becomes gauge invariant without any constraint on gauge field or parameter.

\medskip

In the following subsections, we construct higher-derivative gauge invariant actions making use of these GCS. Our construction essentially follows that of \cite{Francia:2002aa}, although the context here is different.

\subsection{Einstein-like actions}

From the gauge transformations \eqref{Gamma tr}\,,
one can see that multiple $v$-traces of the symbols
\be\label{F2n}
	F^{\sst (s)}_{\sst 2n}(x,u) := (\partial_{v}^{\,2})^{n}\,\Gamma^{\sst (s,2n)}(x;u,v)\,,
\ee
transform into the multiple traces of the gauge parameters
\be
	\delta\,F^{\sst (s)}_{\sst 2n}(x,u)=
	(u\cdot\partial_{x})^{2n+1}\,(\partial_{u}^{\,2})^{n}\,
	\varepsilon^{\sst (s-1)}(x,u)\,.
	\label{F tr}
\ee
For \mt{n=1}\,, the object $F^{\sst (s)}_{\sst 2}$ coincides with the spin-$s$ analogue of the Ricci tensor, given by the Fronsdal's operator $\cF_{\sst 2}$\,
\be
	F^{\sst (s)}_{\sst 2}(x,u)=\cF_{\sst 2}\,\varphi^{\sst (s)}(x,u)\,,\qquad
	\cF_{\sst 2}:=
	\partial_{x}^{\,2}-u\cdot\partial_{x}\,\partial_{u}\cdot\partial_{x}
	+\tfrac1{2}\,(u\cdot\partial_{x})^{2}\,\partial_{u}^{\,2}\,,
	\label{Frons eq}
\ee
and is invariant under the gauge transformations generated by traceless parameters.
For general $n$\,, the $F^{\sst (s)}_{\sst 2n}\,$'s are invariant under gauge transformations with parameters subject to
higher-trace constraints
\be\label{constraint eps}
	(\partial_{u}^{\,2})^{n}\,\varepsilon^{\sst (s-1)}(x,u)=0\,.
\ee
For the subsequent analysis,
we provide another useful expression for $F^{\sst (s)}_{\sst 2n}$\,,
which can be obtained using the identity relating $\Gamma^{\sst (s,r)}$ and $\Gamma^{\sst (s,r-2)}$\,:
\be
	\partial_{v}^{\,2}\,\Gamma^{\sst (s,r)}(x,u,v)
	=\cF_{\sst r}\,\Gamma^{\sst (s,r-2)}(x,u,v)\,.
\ee	
Here we have introduced the generalized Fronsdal operators $\cF_{r}$ as
\ba\label{F r}
	\cF_{\sst r} \edf (\partial_{x}-\tfrac1{r}\,u\cdot\partial_{x}\,\partial_{u})\cdot
	(\partial_{x}-\tfrac1{r-1}\,u\cdot\partial_{x}\,\partial_{u}) \nn
	 \eq \partial_{x}^{\,2}-\tfrac2r\,u\cdot\partial_{x}\,\partial_{u}\cdot\partial_{x}
	+\tfrac1{r(r-1)}\,(u\cdot\partial_{x})^{2}\,\partial_{u}^{\,2}\,.
\ea
After iterations, the Ricci-like tensors $F^{\sst (s)}_{\sst 2n}$\, can be factorized as
\be\label{2deltaequation}
	F^{\sst (s)}_{\sst 2n}(x,u)
	= \cF_{\sst 2n}\,\cF_{\sst 2n-2}\cdots \cF_{\sst 4}\,\cF_{\sst 2}\
	\varphi^{\sst (s)}(x,u)\,,
\ee
where the order of the $\cF_{\sst 2r}$'s is important since they
do not commute with each other.
The $F^{\sst (s)}_{\sst 2n}$\,'s satisfy higher-derivative analogues of the Bianchi-like identities
\be
	\left(\partial_{u}\cdot\partial_{x}
	-\tfrac1{2(n+r)}\,u\cdot\partial_{x}\,\partial_{u}^{\,2}\right)(\partial_{u}^{\,2})^{r}\,
	F^{\sst (s)}_{\sst 2n}=0\,,
	\qquad [r=0,1,\cdots, n-1]\,,
	\label{F recur}
\ee
on the space of the $(n+1)$-th traceless gauge field:
\be\label{constraint f}
    (\partial_{u}^{\,2})^{n+1}\,\varphi^{\sst (s)}(x,u)=0\,.
\ee

Let us now construct the actions giving rise to the equations
\mt{F^{\sst (s)}_{\sst 2n}=0}\,.
In the two-derivative (\mt{n=1}) case, the spin-$s$ Einstein tensor
$G^{\sst (s)}_{\sst 2}$ can be obtained from the Ricci tensor
by a trace modification:
\be
	G^{\sst (s)}_{\sst 2}=\cI_{\sst 2}\,
	F^{\sst (s)}_{\sst 2}\,,
	\qquad
	\cI_{\sst 2}=1-\tfrac14\,u^{2}\,\partial_{u}^{\,2}\,.
	\label{G F}
\ee
Imposing the doubly-traceless constraint on fields,
the Einstein tensor becomes self-adjoint and gives the Fronsdal Lagrangian:
\mt{\mathscr G_{\sst 2}=\ppd{\varphi^{\sst (s)}}{G^{\sst (s)}_{\sst 2}}\,}.
For the higher-derivative ($n\ge2$) cases,
one can also consider the Einstein-like tensors $G^{\sst (s)}_{\sst 2n}$
by appropriately modifying the traces of the Ricci-like
tensors $F^{\sst (s)}_{\sst 2n}$ as
\be\label{Einstein}
 	G^{\sst (s)}_{\sst 2n}=\cI_{\sst 2n}\,F^{\sst (s)}_{\sst 2n}\,,
\ee
where $\cI_{\sst 2n}$ is an operator of form
\mt{\sum_{k} a_{k}\,(u^{2})^{k}\,(\partial_{u}^{2})^{k}}\,.
One can determine $\cI_{\sst 2n}$ either by
requiring the gauge invariance of the action or the self-adjointness of $G^{\sst (s)}_{\sst 2n}$\,.
The gauge invariance requires that the divergence of $G^{\sst (s)}_{\sst 2n}$
gives the Bianchi-like identities \eqref{F recur}, and
this gives the condition
\be\label{A cond}
	\partial_{u}\cdot\partial_{x}\,\cI_{\sst 2n}
	=\tilde\cI_{\sst 2n}\left(\partial_{u}\cdot\partial_{x}
	-\tfrac1{2\,n}\,u\cdot\partial_{x}\,\partial_{u}^{\,2}\right)
\ee
on the operator $\cI_{\sst 2n}$\,,
where $\tilde{\cI}_{\sst 2n}$ is an operator of the same type as $\cI_{\sst 2n}$\,.
The above condition fixes uniquely the operator as
\be\label{I 2n}
	\cI_{\sst 2n}=
	\sum_{r=0}^{[s/2]}\,\frac{1}{r!\,[n]_{r}}\,(-\tfrac14)^{r}\,
	(u^{2})^{r}\,(\partial_{u}^{\,2})^{r}\,,
\ee
where
\be
[a]_{r}:=a(a-1)\cdots(a-r+1)
\ee
are descending Pochhammer symbols.
As one can see from the pole arising when $r=n+1$\,,
the $(n+1)$-th traceless condition \eqref{constraint f}
is indispensable for the Einstein-like actions
\be\label{G Delta F}
\mathscr  G_{\sst 2n}[\varphi^{\sst (s)}]=\ppd{\varphi^{\sst (s)}}{G^{\sst (s)}_{\sst 2n}}\,.
\ee
To recapitulate, these actions are compatible with the gauge fields subject to
the $(n+1)$-th traceless constraint \eqref{constraint f},
and they are  invariant under the gauge transformations generated by
the $n$-th traceless gauge parameters \eqref{constraint eps}.

\subsection{Maxwell-like actions}

Coming back to the construction  \eqref{F2n},
instead of taking only traces one can
also take a divergence, obtaining\footnote{
More generally, one may consider
\be\label{GCF}
	(\pr_x\cdot\pr_v)^m\,(\pr_v^{\,2})^n\,\Gamma^{\sst (s,2n+m)}(x;u,v)\,,
\ee
whose gauge symmetry requires the constraints:
\be
	(\pr_x\cdot\pr_u)^m\,(\pr_u^2)^n\,\varepsilon^{\sst (s-1)}(x,u)=0\,.
\ee
Let us note that when \mt{n=0}\,, the tensor \eqref{GCF} becomes self-adjoint
without any constraint on the field so directly provides the Lagrangian of the theory,
analogously to the \emph{reducible} transverse-invariant theory of \cite{Campoleoni:2012th}
with $m=1$ and the curvature squared theory with $m=s$
which is the local counterpart of \cite{Francia:2010qp}.}
\be
	L^{\sst (s)}_{\sst 2n}(x,u) :=
	\partial_{v}\cdot\partial_{x}\,
	(\partial_{v}^{\,2})^{n-1}\,\Gamma^{\sst (s,2n-1)}(x;u,v)\,,
\ee
then it transforms under the gauge variation as
\be
	\delta\,L^{\sst (s)}_{\sst 2n}(x,u)=
	(u\cdot\partial_{x})^{2n}\,
	\partial_{u}\cdot\partial_{x}\,(\partial_{u}^{\,2})^{n-1}\,	
	\varepsilon^{\sst (s-1)}(x,u)\,.
	\label{F tr}
\ee
For \mt{n=1}\,, the object $L^{\sst (s)}_{\sst 2}$
coincides with the spin-$s$ tensor of transverse-invariant theories
\cite{Skvortsov:2007kz,Campoleoni:2012th},
defined in terms of Maxwell operator $\cL$\,,
\be
	L^{\sst (s)}_{\sst 2}(x,u)=\cL\,\varphi^{\sst (s)}(x,u)\,,\qquad
	\cL:=
	\partial_{x}^{\,2}-u\cdot\partial_{x}\,\partial_{u}\cdot\partial_{x}\,,
	\label{Mx eq}
\ee
and it is invariant under the gauge transformations generated by
transverse parameter.
For general $n$\,, the SV-like tensors $L^{\sst (s)}_{\sst 2n}\,$'s are invariant under
the gauge transformations with the parameters subject to
\be\label{M constraint eps}
	\partial_{u}\cdot\partial_{x}\,(\partial_{u}^{\,2})^{n-1}\,
	\varepsilon^{\sst (s-1)}(x,u)=0\,.
\ee
Moreover, as in the case of $F^{\sst (s)}_{\sst 2n}$,
they can be factorized as
\ba\label{M 2deltaequation}
	L^{\sst (s)}_{\sst 2n}(x,u)
	\eq \cF_{\sst 2n-1}\,\cF_{\sst 2n-3}\cdots \cF_{\sst 5}\,\cF_{\sst 3}\,
	\cL\,
	\varphi^{\sst (s)}(x,u)\nn
	\eq \left(\pr_x^2-\tfrac{1}{2n-1}\,u\cdot\pr_x\,\pr_x\cdot\pr_u\right)
	F_{\sst 2n-2}(x,u)\,.
\ea
From the second expression, one can see that on
the space of $n$-th traceless fields, such that
\be\label{M constraint f}
	(\partial_{u}^{\,2})^{n}\,\varphi^{\sst (s)}(x,u)=0\,,
\ee
they satisfy the Bianchi-like identities:
\ba\label{M bianchiprime}
\left(\pr_x\cdot\pr_u-\tfrac{1}{2(n+r)}\,u\cdot\pr_x\,\pr_u^2\right)
(\partial_{u}^{2})^{r}\,
L_{\sst 2n}(x,u)=0 \qquad [r=0,1,\ldots ,n-1]\,.
\ea
To comply with the $n$-th traceless constraint on the field,
it is necessary to impose
the same condition on the gauge parameter:
\be\label{n+1 tf ep}
   (\partial_{u}^{\,2})^{n}\,\varepsilon^{\sst (s-1)}(x,u)=0\,.
\ee
Finally the action leading to the equation \mt{L^{\sst (s)}_{\sst 2n}=0}
can be determined as
\be
	\mathscr M_{\sst 2n}[\varphi^{\sst (s)}]
	=\ppd{\varphi^{\sst (s)}}{M^{\sst (s)}_{\sst 2n}}\,,
	\qquad
 	M^{\sst (s)}_{\sst 2n}=\cI_{\sst 2n}\,L^{\sst (s)}_{\sst 2n}\,,
\ee
thanks to the Bianchi-like identities \eqref{M bianchiprime}.
To recapitulate, these actions are compatible with gauge fields subject to
the $n$-th traceless constraint \eqref{M constraint f}
and are  invariant under the gauge transformations generated by
parameters subject to the constraints \eqref{M constraint eps} and \eqref{n+1 tf ep}.
One may regard these actions $\mathscr M_{\sst 2n}$
as partially gauge fixed versions of the
Einstein-like actions $\mathscr G_{\sst 2n}$\,, with
gauge fixing \eqref{constraint f}  to \eqref{M constraint f}.
As mentioned in Introduction, the equations
of motion of these theories are shown to be
equivalent to ones involving higher-spin curvature \cite{Francia:2012rg}.

\section{Weyl-like actions}
\label{sec: Weyl}

In the previous section, allowing higher derivatives in the quadratic action we
have constructed the $2n$-derivative Einstein-like and Maxwell-like actions $\mathscr G_{\sst 2n}$
and $\mathscr M_{\sst 2n}$\,.
Starting from \mt{n=1} and increasing the number $n$\,,
there is an enhancement of the gauge symmetries
due to the weakening of the constraints \eqref{constraint eps}
and (\ref{M constraint eps}\,,\,\ref{n+1 tf ep})
imposed on the gauge parameters.
However, when \mt{2n\ge s}, these constraints are completely removed, so that the actions do not acquire any additional symmetry in general.
In other words, the hierarchy of higher-derivative Einstein/Maxwell-like
actions with constrained gauge symmetries covers only half of Table \ref{table}.
It is worth noticing that the \mt{(d+2s-4)}-derivative Einstein- and Maxwell-like actions are exceptions from this point of view. In fact, the trace modifier $\mathcal I_{\sst 2n}$ in $\mathscr G_{\sst 2n}$ and $\mathscr M_{\sst 2n}$ become the trace projector when \mt{2n=d+2s-4}\,:
\be
	\mathcal I_{\sst d+2s-4}\,u^{2}\,\alpha^{\sst (s-2)}=0\,.
\ee
Therefore, these actions acquire additional Weyl symmetry
and coincide with the conformal higher-spin action.

On the other hand, the existence of the $2s$-derivative
higher-spin Weyl action suggests
that there might be another hierarchy of actions with constrained Weyl (and gauge) symmetries.
As discussed in the Introduction, the spin-two Weyl action
can be written as a square of the Einstein tensor.
This gives a crucial hint that Weyl-like actions
can be obtained as products of the Einstein-like or Maxwell-like tensors.
In the following, we present two classes of Weyl-like actions:
 the first one involves $4n$ derivatives while the other $4n+2$ derivatives.

\subsection{$4n$-derivative Weyl-like actions}

\subsubsection*{Four-derivative case}

Let us begin with the linearized Weyl gravity action,
which admits a factorized expression:
\be\label{spin 2 W}
	\mathscr W_{\sst 4}[\varphi^{\sst (2)}]=
	\ppd{G^{\sst (2)}_{\sst 2}}{\left(1-\tfrac1{2(d-1)}\,u^{2}\,\partial_{u}^{2}\right)
	G^{\sst (2)}_{\sst 2}}\,.
\ee
Generalizing the spin-2 Einstein tensor $G^{\sst (2)}_{\sst 2}$ to
the spin-$s$ one $G^{\sst (s)}_{\sst 2}$,
we consider the ansatz:
\be\label{C4}
	\mathscr W_{\sst 4}[\varphi^{\sst (s)}]=\ppd{G^{\sst (s)}_{\sst 2}}{\cA_{\sst 2}\,G^{\sst (s)}_{\sst 2}}\,,
	\qquad
	\cA_{\sst 2}=1+a\,u^2\,\pr_u^2\,,
\ee
for the higher-spin action with Weyl symmetry.
The form of the ansatz through $G^{\sst (s)}_{\sst 2}$
already guarantees the constrained gauge symmetry when implemented
by the Fronsdal constraints (doubly-traceless/traceless constraint on gauge field/parameter).
Turning to the Weyl symmetry, we first notice that its parameter must be traceless,
\be
	\pr_u^{\,2}\,\a^{\sst (s-2)}(x,u)=0\,,
\ee
for compatibility with the doubly-traceless field.
An explicit computation then shows that the Weyl symmetry arises for a special value of $a$ in \eqref{C4},
\be\label{C4a}
	a=-\tfrac1{2(d+2s-5)}\,,
\ee
which generalizes the spin-2 case -- see the coefficient in \eqref{spin 2 W}.
The crucial novelty of the higher-spin case \eqref{C4} with respect to the linearized Weyl gravity \eqref{spin 2 W} is that
the Weyl symmetry is, in fact, constrained with the \emph{transverse}
constraint
\be\label{tta}
	\pr_x\cdot\pr_u\,\a^{\sst (s-2)}(x,u)=0\,.
\ee
Let us remind the reader that
an analogous constraint
has been considered for the \emph{gauge parameter}
in the transverse-invariant theories of
\cite{Skvortsov:2007kz,Campoleoni:2012th}.

To recapitulate, we have found that a four-derivative Weyl-like action does exist
for Fronsdal fields of general spin
with Weyl symmetry parameter subject to differential constraint (\ref{tta}).
One can obtain the same result starting from the Fronsdal constraints
but without assuming the form of the action:
the action \eqref{C4} with a \eqref{C4a} is the unique four-derivative action acquiring Weyl symmetry.
In Section \ref{sec: discussion}, we show that
the spectrum of this action consists of two massless spin $s$
(relatively ghost) and a massless spin $s-1$\, modes,
in analogy with the Weyl gravity.

\subsubsection*{General case}

The form of (\ref{C4}) is suggestive, so that we can now generalize it to higher derivative cases.
Replacing the two-derivative Einstein tensor $G^{\sst (s)}_{\sst 2}$
with the $2n$-derivative one $G^{\sst (s)}_{\sst 2n}$\,,
let us consider the ansatz:
\be\label{Weyl like}
	\mathscr W_{\sst 4n}[\varphi^{\sst (s)}]
	=\ppd{G^{\sst (s)}_{\sst 2n}}{\cA_{\sst 2n}\,G^{\sst (s)}_{\sst 2n}}\,,
\ee
where $\cA_{\sst 2n}$ is the trace modifier to be determined requiring Weyl invariance.
As in the four-derivative case, the constraints imposed on gauge field and parameter follow those of  the Einstein-like action $\mathscr G_{\sst 2n}$\,:
$(n+1)$-th/n-th traceless gauge field/parameter.
Moreover, considering the Weyl symmetry, its parameter
is subject to the $n$-th traceless constraint:
\be
	(\pr_{u}^{2})^{n}\,\a^{\sst (s-2)}(x,u)=0\,,
\ee
for compatibility with the $(n+1)$-th traceless gauge field.

The question we want to turn to is whether the action \eqref{Weyl like},
with an appropriate choice of $\cA_{\sst 2n}$, can admit a Weyl symmetry.
To answer it, we first compute the Weyl variation of the action,
then get an equation for the Weyl parameter $\alpha^{\sst (s-2)}$
that eliminates the variation. This equation defines a constraint for $\alpha^{\sst (s-2)}$ and depends on the form of the operator $\cA_{\sst 2n}$\,.
Hence, the point is whether some operator $\cA_{\sst 2n}$ can lead to
a reasonable constraint on the Weyl parameter $\alpha^{\sst (s-2)}$\,.

Computing the variation of \eqref{Weyl like}
under the $\a^{\sst (s-2)}$ transformation, using \eqref{Einstein}, gives
\be
	\d_{\alpha}\,\mathscr W_{\sst 4n}[\varphi^{\sst (s)}]
	=2\,\ppd{G^{\sst (s)}_{\sst 2n}}{\cC_{\sst 2n}\,
	\d_{\alpha}F^{\sst (s)}_{\sst 2n}}\,,
\ee
where $\cC_{\sst 2n}$ is given by
\be\label{C exp}
	\cC_{\sst 2n}:=\cA_{\sst 2n}\,\cI_{\sst 2n}=
	\sum_{k=0}^{n}\,c_{k}\,(u^{2})^{k}\,(\partial_{u}^{2})^{k}\,.
\ee
Since $\cI_{\sst 2n}$ is fixed and invertible (for \mt{2n\neq d+2s-4}),
determining $\cC_{\sst 2n}$
is equivalent to determining $\cA_{\sst 2n}$\,.
In the following, we first compute the Weyl variation of the Ricci-like tensor
$F_{\sst 2n}^{\sst (s)}$\,,
and then simplify the expression for
$\cC_{\sst 2n}\,\delta_{\alpha}\,F_{\sst 2n}^{\sst (s)}$\,:
\begin{itemize}
\item
For the computation of $\delta_{\alpha}\,F_{\sst 2n}^{\sst (s)}$\,,
we make use of the identities
\ba
	&&\cF_{\sst 2r}\,u^2=u^2\, \cF_{\sst 2r}+\tfrac{d+2\,u\cdot\pr_u-2-4r}{r(2r-1)}\,(u\cdot\pr_x)^2\,,\\
	&&\cF_{\sst 2r}\,(u\cdot\pr_x)^k= \tfrac{(2r-k)(2r-k-1)}{2r(2r-1)}\,(u\cdot\pr_x)^k\,
\cF_{\sst 2r-k}\,,
\ea
together with the Bianchi-like identities (\ref{F recur}).
Employing these,
respectively yields the Weyl variation of the Ricci-like tensors,
\ba\label{ricci var}
	&&\delta_{\alpha} F^{\sst (s)}_{\sst 2n}
	= \cF_{\sst 2n}\,\cdots\,\cF_{\sst 2}\,u^2\,\alpha^{\sst (s-2)} \nn
	&&\quad= \left[-\tfrac{1}{2n-1}\,(u\cdot\pr_x)^2
	\left\{2(n-\t)
	+\tfrac{1}{2(n-1)}\,u^2\,\pr_u^2\right\}
	+u^2\,\pr_x^2\right]\,F_{\sst 2(n-1)}^{\sst (s-2)}(\alpha)\,,
\ea
where \mt{\t=(d+2s-4)/2}
and
\be
F_{\sst 2(n-1)}^{\sst (s-2)}(\alpha):=\cF_{\sst 2(n-1)}\,\cdots\,\cF_{\sst 2}\,\alpha^{\sst (s-2)}.
\ee

\item
Next, we express $\cC_{\sst 2n}\,\delta_{\alpha}\,F_{\sst 2n}^{\sst (s)}$
in the form of a normal ordered operator acting on $F_{\sst 2(n-1)}^{\sst (s-2)}(\alpha)$\,,
using again the Bianchi-like identities (\ref{F recur}) together with the identity:
\ba\label{commutator}
(\pr_u^{\,2})^n(u^2)^m=\sum_{k=0}^{{\rm min}\{n,m\}} &&\!\! 2^{k}\,k!\,
\binom{n}{k}\,\binom{m}{k}\,
\frac{[d+2\,u\cdot\pr_u+2\,(n-m+k-1)]!!}{[d+2\,u\cdot\pr_u+2\,(n-m-1)]!!}\times\nn
&& \times\,(u^2)^{m-k}\,(\pr_u^2)^{n-k}\,,
\ea
which can be proved by induction.
One then finds:
\ba\label{W var}
	\cC_{\sst 2n}\,\d_{\alpha}F^{\sst (s)}_{\sst 2n}
	\eq \sum_{k=0}^{n}\ \frac{n}{(2n-1)(n-k+1)}\times\nn
	&&\qquad \times\,(u^{2})^{k}
	\left[ c^{\sst \sharp}_{k}\,\pr_x^2-
	c^{\sst \flat}_{k}\,(u\cdot\pr_x)^2\,\pr_u^2\right]
	(\pr_u^2)^{k-1}\,F_{\sst 2(n-1)}^{\sst (s-2)}(\alpha)\,,
\ea
where the $c^{\sst \sharp}_{k}$ and $c^{\sst \flat}_{k}$ are coefficients
given in terms of the $c_{k}$'s
\ba\label{c's}
	c^{\sst\sharp}_{k}\eq(2n-2k+1)\,c_{k-1}+4\,k\,(n-k+1)\,(2\t-2k+1)\,c_k\,, \nn
	c^{\sst\flat}_{k}\eq \tfrac{1}{n-k}\left[c_{k-1}+4\,(n-k+1)\,(n+k-\t)\,c_k\right].
\ea
\end{itemize}
The variation \eqref{W var} contains two types of terms:
the first type (with coefficients $c^{\sst\sharp}_{k}$)
does not involve any overall gradient operators, while
the second (with coefficients $c^{\sst\flat}_{k}$)
does involve an overall double gradient operator.
Since this Weyl variation is to be contracted with the Einstein-like tensor
$G^{\sst (s)}_{\sst 2n}$\,,
there is a chance that the terms of the second type vanish
due to the divergence-free nature of $G^{\sst (s)}_{\sst 2n}$ (or equivalently due to the Bianchi-like identities).
On the other hand, the terms of the first type have no chance of vanishing by themselves,
and therefore, we require the $c^{\sst\sharp}_{k}$'s
to vanish, which gives the following recurrence relation
on the coefficients $c_{k}$'s:
\be\label{u less cond}
	c^{\sst\sharp}_{k}=0\quad \Rightarrow \quad
	(2n-2k+1)\,c_{k-1}+4\,k\,(n-k+1)\,(2\t-2k+1)\,c_k=0\,.
\ee
This equation (with the choice \mt{c_0=1}) uniquely determines the trace modifier
$\cC_{\sst 2n}$\,, and consequently $\cA_{\sst 2n}$\,.

After fixing $c^{\sst \sharp}_{k}=0$\,, the Weyl variation
is given by the $c^{\sst\flat}_{k}$ terms with an overall double gradient.
One can integrate by parts one of the gradient operators and get a divergence of
the Einstein-like tensor, $\partial_{u}\cdot\partial_{x}\,G^{\sst (s)}_{\sst 2n}$\,, which
vanishes only when it is contracted with a $n$-th traceless tensor.
Hence, Weyl invariance imposes the condition that
the $n$-th trace of the left-over part (after integrating by part one gradient) be zero:
\be\label{w var=0}
	(\pr_{u}^{2})^{n}\,u\cdot\pr_x\left[\ \sum_{k=0}^{n}\ \frac{c^{\sst \flat}_{k}}{n-k+1}\,
	(u^{2})^{k}\,(\pr_u^2)^{k}\,\right]
	F_{\sst 2(n-1)}^{\sst (s-2)}(\alpha)=0\,.
\ee
Due to the $n$-th traceless constraint on $\a^{\sst (s-2)}$\,,
the above condition reduces to the differential constraint
\be\label{Weyl constraint}
	\pr_u\cdot\pr_x\,(\pr_u^2)^{n-1}\,\cF_{\sst 2(n-1)}\cdots \cF_{\sst 2}\,\alpha^{\sst (s-2)}=0\,.
\ee
For the four-derivative (\mt{n=1}) case, this constraint
reduces to the transversality condition \eqref{tta},
while for the other values of $n$ it gives a rather unusual type of constraint
containing Fronsdal-like operators.
In fact, for the correct analysis of the constraint,
one should take into account the gauge-for-gauge symmetry:
\be\label{gfg}
	\d\, \varepsilon^{\sst (s-1)}(x,u)=u^2\,\b^{\sst (s-3)}(x,u)\,,
	\qquad
	\d\,\a^{\sst (s-2)}(x,u)=-\,u\cdot\pr_x\,\b^{\sst (s-3)}(x,u)\,,
\ee
possessed by the gauge plus Weyl transformations \eqref{gw transf}.
Here $\beta^{\sst (s-3)}$ is the
gauge-for-gauge parameter and satisfies the \mt{(n-1)}-th traceless constraint:
\mt{(\partial_{u}^{\,2})^{n-1}\beta^{\sst (s-3)}=0}\,.
The importance of the gauge-for-gauge transformation \eqref{gfg}
is that using it one can always make the Weyl parameter $\alpha^{\sst (s-2)}$
traceless:
\be\label{gfg t a}
	\pr_u^{\,2}\, \a^{\sst (s-2)}(x,u)=0\,.
\ee
In other words, the trace part of Weyl transformation can always be
expressed as a \emph{gauge} transformation.
Let us notice also that eq. \eqref{Weyl constraint} is invariant under
the transformation  \eqref{gfg} for $\alpha^{\sst (s-2)}$\,.
Taking into account the condition \eqref{gfg t a}, the unusual constraint \eqref{Weyl constraint}
reduces to the requirement of vanishing multiple divergence:
\be\label{2n-1div}
	(\pr_x\cdot\pr_u)^{2n-1}\,\a^{\sst (s-2)}(x,u)=0\,.
\ee

\subsection{$(4n+2)$-derivative Weyl-like actions}
\label{W4N+2}

\subsubsection*{Six-derivative case}

Let us begin with the six-derivative spin-3 Weyl action
(see Appendix \ref{sec: s3 Weyl}).
Differently from linearized Weyl gravity, it is not given
as an Einstein tensor squared but admits another type of factorization:
\be\label{6 der W}
	\mathscr W_{\sst 6}[\varphi^{\sst (3)}]=
	\ppd{F^{\sst (3)}_{\sst 2}}{
	\left[\pr_x^2
	-\tfrac{d-2}{6(d+1)}\,u\cdot\pr_x\,\pr_x\cdot\pr_u
	-\tfrac{d+4}{12(d+1)}\,u^2\,\pr_x^{\,2}\,\pr_u^{\,2}\right]
	\,F^{\sst (3)}_{\sst 2}}\,.
\ee
Generalizing the spin-3 Ricci tensor $F^{\sst (3)}_{\sst 2}$ to
the spin-$s$ one  $F^{\sst (s)}_{\sst 2}$\,,
we consider for the Weyl invariant action the ansatz
\be\label{W6}
	\mathscr W_{\sst 6}[\varphi^{\sst (s)}]=
	\ppd{F^{\sst (s)}_{\sst 2}}
	{\left[\pr_x^2+a\,u\cdot\pr_x\,\pr_x\cdot\pr_u+ b\,u^2\,\pr_x^{\,2}\,
	\pr_u^{\,2}\right] F^{\sst (s)}_{\sst 2}}\,,
\ee
where $\varphi^{\sst (s)}$ is  doubly traceless.
Note that this ansatz is the most general one satisfying
\emph{i)} manifest self-adjoint-ness and
\emph{ii)} factorization in terms of two $F^{\sst (s)}_{\sst 2}$'s.
More precisely, there can be other two-derivative operators inside
of the square bracket in \eqref{W6},
but they can be all replaced with the actual ones
by the virtue of Bianchi-like identities.

The traceless gauge symmetry of \eqref{W6} is ensured by the presence of the Fronsdal tensor $F^{\sst (s)}_{\sst 2}$.
On the other hand, as one can check by explicit computations,
the Weyl symmetry arises only for
\be\label{a&b}
	a=-\tfrac{d+2s-8}{6(d+2s-5)}\,,\qquad
	b=-\tfrac{d+2s-2}{12(d+2s-5)}\,,
\ee
with a parameter subject to the traceless and doubly transverse constraints:
\be
	\pr_u^{\,2}\,\a^{\sst (s-2)}(x,u)=0\,,
	\qquad  (\pr_x\cdot\pr_u)^2\,\a^{\sst (s-2)}(x,u)=0\,.
\ee
Notice that the above constraints coincide with the formal 6-derivative interpolation of
the previously found $4n$-derivative constraints (\ref{gfg t a}\,,\,\ref{2n-1div}), obtained for $n=\tfrac{3}{2}$.

Turning back to the expression \eqref{W6},
one may wonder whether it can be recast into a simpler form
as in the four-derivative case.
Given that it involves six-derivatives, the action cannot be written as a square of
Einstein-like or Maxwell-like tensors,
but it can, in fact, be expressed as a product between
Einstein and four-derivative Maxwell-like tensors as
\be\label{W6n}
	\mathscr W_{\sst 6}[\varphi^{\sst (s)}]=
	\ppd{G^{\sst (s)}_{\sst 2}}
	{\left(1-\tfrac{1}{2(d+2s-5)}\,u^2\,\pr_u^{\,2}\right) M^{\sst (s)}_{\sst 4}}\,.
\ee
Let us notice that the trace modifier lying between the Einstein and Maxwell-like
tensors coincides with  that of the four-derivative action -- see \eqref{C4a}.
Moreover, the expression \eqref{W6n} shows clearly
that the action actually admits a larger gauge symmetry:
that of the four-derivative Maxwell-like action $\mathscr M_{\sst 4}$
(\ref{M constraint eps}\,,\,\ref{n+1 tf ep}) rather
than the Fronsdal one with traceless parameter.

\subsection*{General case}

The \mt{(4n+2)}-derivative Weyl-like action
can be obtained
generalizing the six-derivative one \eqref{W6n} to
\be
\label{C4N+2}
	\mathscr W_{\sst 4n+2}[\varphi^{\sst (s)}]
	=\ppd{G^{\sst (s)}_{\sst 2n}}{\cA_{\sst 2n}\,
	M^{\sst (s)}_{\sst 2n+2}}\,,
\ee
where the gauge field is subject to the \mt{(n+1)}-th traceless constraint.
The kinetic operator in the above action is self-adjoint:
first, the Maxwell-like tensor can be written as
\mt{M^{\sst (s)}_{\sst 2n+2}=\cK_{\sst 2}\,G_{\sst 2n}^{\sst (s)}} with a
two-derivative operator $\cK_{\sst 2}$\,,
then $\cA_{\sst 2n}\,\cK_{\sst 2}$ can be recast into a manifestly
self-adjoint form using the Bianchi-like identities \eqref{F recur}.
From the self-adjointness, one can see
that the gauge symmetry of \eqref{C4N+2} is that of the $(2n+2)$-derivative
Maxwell-like action $\mathscr M_{\sst 2n+2}$\,.
Moreover, the Weyl variation of \eqref{C4N+2} reads simply
\be
	\delta_{\alpha}\,\mathscr W_{\sst 4n+2}[\varphi^{\sst (s)}]
	=2\,\ppd{\delta_{\alpha}\,G^{\sst (s)}_{\sst 2n}}{\cA_{\sst 2n}\,
	M^{\sst (s)}_{\sst 2n+2}}\,,
\ee
so that the analysis goes along the same lines as in the $4n$-derivative case:
requiring the absence of $c^{\sst\sharp}_{k}$ terms in
the Weyl variation \eqref{W var},
the operator $\cA_{\sst 2n}$
is completely determined  by \eqref{u less cond}.
The remained $c^{\sst\flat}_{k}$ terms are
proportional to double gradients,
so that integrating by parts one gradient one gets a divergence of
Maxwell-like tensor, \mt{\partial_{u}\cdot\partial_{x}\,M^{\sst (s)}_{\sst 2n+2}}\,.
The latter vanishes (differently from the $4n$-derivative case where one gets
\mt{\partial_{u}\cdot\partial_{x}\,G^{\sst (s)}_{\sst 2n}})
when it is contracted
with a tensor whose divergence of the $n$-th trace vanishes
-- see \eqref{M constraint eps}, so that the constraint on the Weyl parameter finally reads
\be\label{Weyl constraint2}
	(\pr_u\cdot\pr_x)^2\,(\pr_u^2)^{n-1}\,\cF_{\sst 2(n-1)}\cdots \cF_{\sst 2}\,\alpha^{\sst (s-2)}=0\,.
\ee
After using the gauge-for-gauge freedom to reach the traceless condition \eqref{gfg t a},
one gets the multiple divergence constraint
\be\label{Wc4N+2}
(\pr_x\cdot\pr_u)^{2n}\,\a^{\sst (s-2)}(x,u)=0\,.
\ee
The Skvortsov-Vasiliev action \cite{Skvortsov:2007kz} is a member of this hierarchy with $n=0$.

\subsection{Emergence of Francia's mass term}

In linearized Weyl gravity (or more generally in the four-derivative Weyl-like actions),
the trace modifier appearing between two Einstein tensors turns out to coincide
with the inverse of the Fierz-Pauli mass term:
\be\label{M2}
	(\cA_{\sst 2})^{-1}=\cI_{\sst\rm FP}=1-\tfrac{1}{2}\,u^2\,\pr_u^{\,2}\,.
\ee
Therefore, one may expect that the trace modifiers $\cA_{\sst 2n}$'s
appearing in the Weyl-like actions
be also inverses of some mass operators.
In order to check this idea, one may explicitly
compute the inverse of $\cA_{\sst 2n}$\,:
\be
	(\cA_{\sst 2n})^{-1}=\cI_{\sst 2n}\,(\cC_{\sst 2n})^{-1}\,.
\ee
However, there is a shortcut.
Instead of directly computing $(\cA_{\sst 2n})^{-1}$\,,
we first conjecture that they all coincide with the higher-spin
analogue of Fierz-Pauli mass term
introduced by Francia in \cite{Francia:2007ee},
\be\label{M fran}
	\cI_{\sst\rm F}=1-\tfrac12\,u^{2}\,\pr_{u}^{2}
	-\tfrac1{8}\,(u^{2})^{2}\,(\pr_{u}^{2})^{2}
	-\cdots-\tfrac1{2^{k}\,k!\,(2k-3)!!}\,(u^{2})^{k}\,(\pr_{u}^{2})^{k}-\cdots\,,
\ee
which is determined by the property:
\be\label{m condition}
	\pr_x\cdot\pr_u\,\cI_{\sst\rm F}
	=\tilde\cI_{\sst\rm F}\,\left(\pr_x\cdot\pr_u-u\cdot\pr_x\,\pr_u^2\right)\,.
\ee
Here $\tilde\cI_{\sst\rm F}$ is some operator of the same type as $\cI_{\sst\rm F}$\,.
Rewriting the relation to the mass operator in a different way
\be
	\cI_{\sst\rm F}=\cI_{\sst 2n}\,(\cC_{\sst 2n})^{-1}\quad
	\Leftrightarrow \quad \cI_{\sst\rm F}\,\cC_{\sst 2n}=\cI_{\sst 2n}\,,
\ee
the property \eqref{m condition} of $\cI_{\sst\rm F}$, together
with the property \eqref{A cond}\footnote{
From the conditions \eqref{A cond} and \eqref{m condition},
one can see that the Francia mass operator actually
belongs to the class of the trace modifiers $\cI_{n}$ as
$\cI_{\sst\rm F}=\cI_{\sst 1}$\,.}
of $\cI_{\sst 2n}$,
induces a condition on the operator $\cC_{\sst 2n}$ of \eqref{C exp}\,:
\be\label{C cond}
	\left(\pr_x\cdot\pr_u-u\cdot\pr_x\,\pr_u^2\right)\cC_{\sst 2n}
	=\tilde \cC_{\sst 2n}\left(\pr_x\cdot\pr_u-\tfrac1{2n}\,u\cdot\pr_x\,\pr_u^2\right),
\ee
where $\tilde{\cC}_{\sst 2n}$ is again some operator of the same type as $\cC_{\sst 2n}$\,.
This condition determines completely the operator $\cC_{\sst 2n}$
with $c_{0}=1$\,.
To see how this works, let us first consider the relations
\ba\label{C con id}
	&& \pr_x\cdot\pr_u\,\cC_{\sst 2n}=\sum_{k=0}^{\infty}\,
	(u^2)^{k-1}\,(\pr_u^2)^{k-1}\,\Big\{
	2\,k\,c_{k}\,u\cdot\pr_{x}\,\pr_{u}^{\,2}
	+\big[c_{k-1}+4\,k\,(k-1)\,c_{k}\big]\,\pr_{x}\cdot\pr_{u}
	\Big\}\,,\nn
	&& u\cdot\pr_{x}\,\pr_u^{\,2}\,\cC_{\sst 2n}=\sum_{k=0}^{\infty}\,
	(u^2)^{k-1}\,(\pr_u^{\,2})^{k-1}\,
	\big[c_{k-1}+2\,k \left(d+2\,u\cdot\pr_u-2k\right)\,c_k\,\big]\times \nn
	&& \hspace{155pt}\times\,
	\big[u\cdot\pr_{x}\,\pr_{u}^{\,2}-2(k-1)\,\pr_{x}\cdot\pr_{u}\big]\,,
\ea
where the $c_{k}$'s are the coefficients appearing in the expansion of $\cC_{\sst 2n}$ \eqref{C exp}.
The left-hand side of the difference between two equations in \eqref{C con id}
coincides with \eqref{C cond}.
Focusing on the right-hand side, the terms proportional to $\pr_{x}\cdot\pr_{u}$
and $u\cdot\pr_{x}\,\pr_{u}^{\,2}$ give respectively
\ba
	&\pr_{x}\cdot\pr_{u}\quad \Rightarrow \quad &
	\tilde c_{k-1}=(2k-1)\,c_{k-1}-
	4\,k\,(k-1)\,(d+2s-2k-3)\,c_{k}\,, \nn
	&u\cdot\pr_{x}\,\pr_{u}^{\,2} \quad \Rightarrow \quad
	& \tfrac1{2n}\,\tilde c_{k-1}= c_{k-1}+2\,k\,(d+2s-2k-3)\,c_{k}\,,
\ea
where $\tilde c_{k}$'s are the coefficients of $\tilde \cC_{\sst 2n}$\,.
Solving for the $\tilde c_{k}$'s from the above equations,
one ends up with a recurrence relation between $c_{k-1}$ and $c_{k}$\,,
which exactly coincides with \eqref{u less cond}.
This proves the conjecture \mt{(\cA_{\sst 2n})^{-1}=\cI_{\sst\rm F}}\,.
Let us comment here that the tensors $S^{\sst (s)}_{\sst 2n}:=
\mathcal I_{\rm\sst F}^{\,-1}G^{\sst (s)}_{\sst 2n}$ can be regarded
as generalizations of Schouten tensor in the sense that they transform
under the Weyl transformation as double gradient:
\mt{\delta_{\alpha}\,S^{\sst (s)}_{\sst 2n}=(u\cdot\partial_{x})^{2}\,(\,\cdots)\,}.

\section{Discussion}
\label{sec: discussion}

In this paper, we have constructed higher-derivative actions for higher spins
which are gauge and Weyl invariant with some constraints.
For a given spin $s$\,, we have first considered
 Einstein-like and Maxwell-like actions involving from 2 to $s$ derivatives.
 These actions proved essential for the construction of Weyl-like actions.
 The latter are associated with \mt{(n+1)}-th traceless gauge fields, such that
\mt{(\pr_{u}^{2})^{n+1}\,\varphi^{\sst (s)}=0}\,,
and consist of two classes:
\begin{itemize}
\item
The first is the $4n$-derivative one given by
\ba\label{C4N}
	\mathscr W_{\sst 4n}[\varphi^{\sst (s)}]
	\eq \ppd{G^{\sst (s)}_{\sst 2n}}{\cI_{\sst\rm F}^{\,-1}\,G^{\sst (s)}_{\sst 2n}} \nn
	\eq \ppd{\varphi^{\sst (s)}}
	{\cF_{\sst 2}^{\dagger}\cdots \cF_{\sst 2n}^{\dagger}\,\cI_{\sst 2n}\,
	\cI_{\sst\rm F}^{\,-1}\,
	\cI_{\sst 2n}\,\cF_{\sst 2n}\cdots \cF_{\sst 2}\,\varphi^{\sst (s)}}\,,
\ea
where $\mathcal F_{\sst n}$\,, $\mathcal I_{\sst 2n}$ and $\mathcal I_{\sst\rm F}$
are given in \eqref{F r}\,, \eqref{I 2n} and \eqref{M fran}.
This action is invariant under the gauge and Weyl transformations \eqref{gw transf} with
\be\label{dc4N}
	(\pr_{u}^{\,2})^{n}\,\varepsilon^{\sst (s-1)}=0\,,\qquad
	 \pr_{u}^{\,2}\,\a^{\sst (s-2)}=0
	 =(\pr_u\cdot\pr_x)^{2n-1}\,\alpha^{\sst (s-2)}\,.
\ee
\item
The second is the \mt{(4n+2)}-derivative one given by
\ba\label{W4N+2}
	\mathscr W_{\sst 4n+2}[\varphi^{\sst (s)}]
	\eq \ppd{G^{\sst (s)}_{\sst 2n}}{\cI_{\sst\rm F}^{\,-1}\,
	M^{\sst (s)}_{\sst 2n+2}} \nn
	\eq \ppd{\varphi^{\sst (s)}}
	{\cF_{\sst 2}^{\dagger}\cdots \cF_{\sst 2n}^{\dagger}\,\cI_{\sst 2n}\,
	\cI_{\sst\rm F}^{\,-1}\,
	\cI_{\sst 2n+2}\,\cF_{\sst 2n+1}\cdots \cF_{\sst 1}\,\cL\,\varphi^{\sst (s)}}\,,
\ea
which is invariant under the gauge and Weyl transformations \eqref{gw transf} with
\be\label{dc4N}
	(\pr_{u}^{\,2})^{n+1}\,\varepsilon^{\sst (s-1)}=0
	=\partial_{x}\cdot\partial_{u}\,(\pr_{u}^{\,2})^{n}\,\varepsilon^{\sst (s-1)}\,,
	\qquad
	 \pr_{u}^{\,2}\,\a^{\sst (s-2)}=0
	 =(\pr_u\cdot\pr_x)^{2n}\,\alpha^{\sst (s-2)}\,.
\ee
\end{itemize}
Notice that for $s=2n$ or $2n+1$\,,
all constraints  are dropped,
and the action $\mathscr W_{\sst 2s}[\varphi^{\sst (s)}]$ coincides with
the higher-spin Weyl action.

The Weyl-like actions $\mathscr W_{\sst 2n}$\,, being higher-derivative, contain ghost modes
in the spectrum and, as a result, lead to non-unitary representations of the Poincar\'e algebra.
Although non-unitary, however, these may still exhibit interesting mathematical properties.
Let us recall that the spin-two Weyl action propagates,
around a flat background, two (relatively ghost) helicity two modes and a helicity one
mode \cite{Riegert:1984hf}.
Interestingly, analyzed around an (A)dS background, the above spectrum
groups into two packages:
a massless spin two
and a partially-massless spin two \cite{Maldacena:2011mk,Deser:2012qg}.
Diagrammatically, this can be expressed as
\be
\mathscr W_{\sst 4}[\varphi^{\sst (2)}]\quad \Rightarrow \qquad
  \begin{array}{c} +\\2 \\ \phantom{1}\end{array}   \quad
 \begin{array}{*1{c}}
    -\\ \tikzmark{left}{2} \\ \tikzmark{right}{1}
   \end{array}
\ee \Highlight
\!\!where each number indicates the corresponding helicity mode
while the enclosure means that the helicities therein become an irreducible set in (A)dS as pertains to a partially-massless representation \cite{Deser:2001us}.
Moreover, the analysis of the spectrum for spin-$s$ Weyl action around
a flat background shows
 that it propagates $\ell$ copies of helicity-$\ell$ modes with \mt{\ell=s, s-1, \ldots, 1}
\cite{Metsaev:2007rw}.
From the analogy of the spin two case, it is natural to expect that,
when deformed to an (A)dS background, all these spectra
group into  partially-massless spin-$s$ modes
with alternating signs of their kinetic operators.
As in the spin two case, this can be summarized via the following diagram
\be
\mathscr W_{\sst 2s}[\varphi^{\sst (s)}]\quad \Rightarrow \qquad
  \begin{array}{c} + \\ s\\  \\ \phantom{\vdots}  \\  \\ \phantom{1} \end{array}
  \quad\quad
  \begin{array}{c} - \\ \tikzmark{left}{$\phantom{s}s\phantom{s}$} \\
	\tikzmark{right}{$s\!-\!1$\!}  \\
	\phantom{\vdots} \\  \\ \phantom{1} \end{array}
  \Highlight \ \ \quad
  \ldots \quad\
   \begin{array}{c}  \pm \\ \tikzmark{left}{$\phantom{s}s\phantom{s}$}
   \\ s\!-\!1 \\ \vdots \\ \tikzmark{right}{\phantom{s}2\phantom{s}}
    \\ \phantom{1} \end{array}
   \Highlight \quad
 \begin{array}{c}  \mp \\ \tikzmark{left}{ $\phantom{s}s\phantom{s}$ } \\ s\!-\!1  \\ \vdots \\ 2 \\
 \tikzmark{right}{\phantom{s}1\phantom{s}}\end{array}
   \Highlight
\ee
where the $r$-th block corresponds to partially-massless spin $s$
of the $r$-th point (or, equivalently, of depth $r$).
While each block is irreducible under the (A)dS isometry group,
the entire spectrum provides a indecomposable
representation of the conformal group \cite{Metsaev:2007rw}.
This gives a hint for a novel class of non-unitary
but interesting representations of the conformal group,
covering all short representations of the isometry group.

One may  expect that there exist even a \mt{(2r+2)}-derivative action
propagating partially-massless fields from the zero-th  (massless) point
to the $r$-th point.
\be\label{spec Weyl-like}
  \begin{array}{c}  +\\ s\\  \\ \phantom{\vdots}  \\  \phantom{1} \end{array}
  \quad \quad
  \begin{array}{c} - \\ \tikzmark{left}{$\phantom{s}s\phantom{s}$} \\
	\tikzmark{right}{$s\!-\!1$\!}  \\
	\phantom{\vdots} \\  \phantom{1} \end{array}
  \Highlight \ \ \quad
    \ldots \quad
  \begin{array}{c} \pm\\  \tikzmark{left}{$\phantom{s}s\phantom{s}$}
   \\ s\!-\!1 \\ \vdots \\ \tikzmark{right}{$s\!-\!r$\!} \end{array}
   \Highlight
\ee
If this action exists, the (A)dS deformation of the Weyl-like action
$\mathscr W_{2r+2}$ can be a good candidate.
In fact, it is the case for \mt{r=1}\,: the four-derivative Weyl-like action
around (A)dS can be obtained as
\be
	\mathscr W^{\L}_{\sst 4}[\varphi^{\sst (s)}]=
	\ppd{G_{\sst 2}^{\L\sst(s)}}
	{\cI_{\sst\rm F}^{\,-1}\,
	G^{\L\sst (s)}_{\sst2}}
	-\eta\,\L\,
	\ppd{\varphi^{\sst (s)}}{G^{\L\sst (s)}_{\sst 2}}
	\qquad \Big[\eta=\tfrac{2(d+2s-6)}{(d-1)(d-2)}\Big]\,,
\ee
adding a Fronsdal action to the four-derivative part.
Here, $G^{\L\sst(s)}_{\sst 2}=\cI_{\sst 2}\,\cF^{\L}_{\sst 2}\,\varphi^{\sst (s)}$
is the spin-$s$ \emph{cosmological} Einstein tensor,
and the field $\varphi^{\sst (s)}(x,u)$
is contracted  with flat auxiliary variables
$u^{\alpha}$'s and the AdS vielbein $\bar e_{\alpha}^{\ \mu}$  as
\be
    \varphi(x,u) = \tfrac{1}{s!}\,u^{\alpha_1}\,\cdots\, u^{\alpha_s}\
    \bar e_{\alpha_1}^{\ \,\mu_1}(x)\, \cdots\, \bar e_{\alpha_s}^{\ \,\mu_s}(x)\
    \varphi_{\mu_1\cdots \mu_s}(x)\,.
\ee
The Fronsdal operator $\mathcal{F}_{\sst 2}^{\Lambda}$ in (A)dS  is given,
in terms of the covariant derivative
\be
D_{\alpha} = \bar e_{\alpha}^{\ \,\mu}\, \nabla_{\mu}
    + \tfrac12\,\bar \omega^{\ \ \, \gamma}_{\alpha\beta}\,
    u^{\b}\, {\partial_{u^{\g}}},
\ee
by
\be
    \mathcal{F}_{\sst 2}^{\Lambda}=
    D^{2}-u\cdot D\,
    \partial_{u}\cdot D
    +\tfrac12\,(u\cdot D)^{2}\,\partial_{u}^{2}
    +\tfrac{2\,\L}{(d-1)(d-2)}\left[u^{2}\,\partial_{u}^{\,2}
    +s^{2}+(d-6)\,s-2(d-3)\right].
\ee
Similarly to the spin two case, this action can be recast in the form:
\be\label{W4 as m pm}
	\mathscr W^{\L}_{\sst 4}=
	\eta\,\Lambda\left[\,-\ppd{\varphi^{\sst (s)}}{G^{\L}_{\sst 2}(\varphi^{\sst (s)})}
	+\ppd{\chi^{\sst (s)}}{
	G^{\L}_{\sst 2}(\chi^{\sst (s)})-\eta\,\L\,\cI_{\sst\rm F}\,\chi^{\sst (s)}}\right],
\ee
where the first term corresponds to the spin-$s$ Fronsdal action
while the second describes the partially-massless spin $s$ of the first point:
the latter admits a gauge description via doubly-traceless tensors
$\chi^{\sst (s)}$ and $\chi^{\sst (s-1)}$\,,
with corresponding traceless gauge parameters
$\alpha^{\sst (s-1)}$ and $\alpha^{\sst (s-2)}$\,.
After gauge fixing the Stueckelberg field $\chi^{\sst (s-1)}$
using $\alpha^{\sst (s-1)}$ and \mt{\partial_{u}\cdot D\,\alpha^{\sst (s-2)}}\,,
one ends up with the system in \eqref{W4 as m pm} with
the residual gauge symmetry:
\be\label{gs pm 1}
	\delta_{\alpha}\,\chi^{\sst (s)}=
	\left[(u\cdot D)^{2}+\L\,u^{2}\right]\alpha^{\sst (s-2)}\,,
	\qquad
	\partial_{u}\cdot D\,\alpha^{\sst (s-2)}=0=
	\partial_{u}^{\,2}\,\alpha^{\sst (s-2)}\,.
\ee
For more details, see e.g. \cite{Deser:2001us,Zinoviev:2001dt,Metsaev:2009hp}.
Let us notice that
the transversality constraint on the Weyl parameter $\alpha^{\sst (s-2)}$
of $\mathscr W^{\L}_{\sst 4}$ arises by a partial gauge fixing procedure,
as the transversality constraints on $\varepsilon^{\sst (s-1)}$ of
$\mathscr M_{\sst 2n}$
(which can be obtained by partially gauge fixing $\mathscr G_{\sst 2n}$).
Hence, there may exist an action
without any differential constraint on the Weyl parameter but
involving auxiliary fields,
such that gauge fixing all the auxiliary fields leads to $\mathscr W_{\sst 2n}$\,.
Such an action may have an ordinary derivative formulation, involving an off-shell field for each propagating degree of freedom, analogous to those considered in \cite{Metsaev:2007rw,Metsaev:2009hp}.

The problem of non-unitarity in these theories
can be in principle handled as in the gravity case:
\begin{itemize}
\item
In AdS background,
one can select only the massless spin $s$, as in \cite{Maldacena:2011mk}, making use of suitable boundary conditions.
\item
In three dimensions,
ghosts do not propagate assuming
the spectrum \eqref{spec Weyl-like}:
Weyl action $\mathscr{W}_{\sst 2s}$ propagates a single scalar mode, while the other members do not have any propagating content.
Hence, it would be interesting to consider these actions
in the context of AdS$_{3}$/CFT$_{2}$ correspondence.
%
\end{itemize}

To conclude, let us discuss briefly the generalization of our actions
to interacting ones.
Despite many efforts, no deformation of AdS Fronsdal action to fully interacting one
is available, while Vasiliev's
equations \cite{Vasiliev:1990en,Vasiliev:2004cp} describe propagation of
an infinite tower of massless interacting higher spins.
On the other hand, the linear conformal higher-spin action ($d=4$ Weyl action)
can be deformed into a fully non-linear action\footnote{
The conformal higher-spin action, analogously to the gravity case,
 might make use of non-linear Weyl tensor
-- (appropriately defined) traceless part  of non-linear higher-spin curvature.
See \cite{Manvelyan:2010jf} for an attempt on the non-linear deformation of
the curvature.
See also \cite{Fradkin:1989md} for the cubic interaction in the frame-like approach.} without
auxiliary fields \cite{Segal:2002gd,Bekaert:2010ky} but with the entire tower of higher-spin fields.
From the conformal gravity,
one can obtain the Einstein action at the level of interacting theory \cite{Metsaev:2007fq,Maldacena:2011mk}.\footnote{
On the other hand, it has been shown that extracting the other component of conformal gravity -- partially-massless spin two -- faces a consistency problem associated with the interaction structure \cite{Deser:2012qg}.}
Hopefully, a similar mechanism can also work for the higher-spin case,
providing new insights for a more conventional action principle\footnote{
See also the recent proposal \cite{Boulanger:2011dd} and references therein.}
 leading to Vasiliev's equations.

\acknowledgments

We would like to thank Luca Lopez, Massimo Taronna, Andrew Waldron and especially Dario Francia and Augusto Sagnotti for fruitful discussions and useful comments.
This work was supported in part by Scuola Normale Superiore, by INFN (I.S. TV12) and by the MIUR-PRIN contract 2009-KHZKRX.
The work of KM is also supported
by the ERC Advanced Investigator Grants no. 226455 ``Supersymmetry,
Quantum Gravity and Gauge Fields'' (SUPERFIELDS).

\appendix

\section{Spin three Weyl action}
\label{sec: s3 Weyl}

In this section, we construct spin three Weyl action in a form
which is suitable for the generalization to six-derivative
Weyl-like action for any spin.

The variation of the spin three Fronsdal tensor
with respect to gauge transformation
$\d \varphi_{\mu\nu\rho}=3\,\pr_{(\mu}\varepsilon_{\nu\rho)}$ gives
\be
	\d {F}_{\mu\nu\rho}=3\,\pr_{\mu}\pr_{\nu}\pr_{\rho}\,
	\varepsilon_{\sigma}^{\ \,\sigma}\,,
\ee
and therefore the following tensor (antisymmetric with respect to first two indices and symmetric with respect to second two):
\be
C_{\m\n,\r\s}=\pr_{\m}F_{\n\rho\sigma}-\pr_{\n}{F}_{\m\rho\sigma}
\ee
is gauge invariant.
The spin three Weyl Lagrangian can be conveniently expressed
in terms  of $C_{\m\n,\r\s}$ as
\be
\mathcal{L}=-\tfrac{1}{2}\,C_{\m\n,\r\s}\,C^{\m\n,\r\s}+\tfrac{d+4}{8(d+1)}\,
C^{'}_{\m\n}\,C^{'\m\n}\qquad
[\,C^{'}_{\m\n}=C_{\m\n,\r}^{\ \ \ \ \, \r}\,]\,.
\ee
Up to integration by parts,
the above Lagrangian can be recast into this form
\be
\mathcal{L}={F}_{\m\n\r}\,\Box\,{F}^{\m\n\r}-\tfrac{d+4}{2(d+1)}\,
{F}^{'}_{\m}\,\Box\,{F}^{'\m}
-\tfrac{d-2}{2(d+1)}\,{F}^{\m\n\r}\,\pr_{(\m}\pr^{\s}{F}_{\n\r)\s}
\qquad
[\,{F}^{'}_{\m}={F}_{\m\n}^{\ \ \, \n}\,]\,,
\ee
which has been used in \eqref{6 der W}  for generalization to any spin.

\section{Spectrum of the four-derivative gauge invariant action}
\label{sec: spec}

In Section \ref{sec: discussion}, we analyzed the spectrum of
the four-derivative Weyl-like action.
To complete the study of the spectrum of four-derivative actions
considered in this paper, we derive here the spectrum of the
four-derivative Einstein-like action.
Since the Maxwell-like actions can be obtained by partially gauge fixing
the Einstein-like ones, the spectrum of the two should coincide
(as in the case of SV and Fronsdal actions).

The four-derivative Einstein-like action $\mathscr G_{\sst 4}$
gives the equation of motion:
\be
	0=F^{\sst (s)}_{\sst 4}
	:=\cF_{\sst 4}\,\cF_{\sst 2}\,\varphi^{\sst (s)} =\left[(\partial_{x}^{\,2})^{2}
	-u\cdot\pr_x\,\cD\right] \varphi^{\sst (s)}\,,
\ee
where
\be
	\cD=\partial_{x}^{\,2}\,\pr_x\!\cdot\pr_u
	-\tfrac{1}{6}\,u\cdot\pr_x
	\left[\partial_{x}^{\,2}\,\pr_u^{\,2}+2\,(\pr_x\!\cdot\pr_u)^2\right]
	+\tfrac{1}{6}\,(u\cdot\pr_x)^2\,\pr_x\!\cdot\pr_u\,\pr_u^{\,2}
	-\tfrac{1}{24}\,(u\cdot\pr_x)^{3}\,(\pr_u^{\,2})^{2}\,.
\ee
The three-derivative operator $\cD$ satisfies
the following properties:
\be
	(\pr_u^{\,2})^2\,\cD\,\varphi^{\sst (s)}=0\,,
	\qquad
	\delta_{\varepsilon} \left(\cD\,\varphi^{\sst (s)}\right)
	=(\partial_{x}^{\,2})^{2}\,\varepsilon^{\sst (s-1)}\,,	
\ee
analogously to the de Donder operator in Fronsdal's theory.
In particular, the second property implies that the gauge condition
\be\label{ga con}
	\cD\,\varphi^{\sst (s)}=0\,,
\ee
is an allowed one.
This gauge condition gives for the gauge field and the parameter (associated with the residual gauge symmetry) the equations
\be\label{eqnb}
	(\partial_{x}^{\,2})^{2}\,\varphi^{\sst (s)}=0\,,\qquad
	(\partial_{x}^{\,2})^{2}\,\varepsilon^{\sst (s-1)}=0\,,
\ee
which are subject to the trace constraints
\be\label{tr constr}
	(\pr_u^{\,2})^3\,\varphi^{\sst (s)}=0\,,
	\qquad
	(\pr_{u}^{\,2})^{2}\,\varepsilon^{\sst (s-1)}=0\,.
\ee
Now we need first to solve these equations and
fix the residual gauge symmetries.
We can then plug back the solutions into \eqref{ga con}
to obtain the on-shell constraints.
Solving the latter will yield the spectrum of the theory.

The solution of the gauge fixed equation (\ref{eqnb}) is
\be\label{solutionb}
	\varphi^{\sst (s)}(x,u)=\int d^{d}p\ \delta(p^{2})
	\left[ \tilde\varphi_{\sst1}^{\sst (s)}(p,u)
	+n\cdot x\,\tilde\varphi_{\sst 2}^{\sst (s)}(p,u)\right] e^{ip\cdot x}\,,
\ee
where $n^\m$ is a time-like vector which we choose \mt{n^{\mu}=\delta^{\mu}_{0}}\,. The Fourier mode $\tilde \varphi_{\sst1}^{\sst (s)}$ corresponds to the regular solutions, while $\tilde\varphi_{\sst 2}^{\sst (s)}$ to the ghost ones.
After a suitable Lorentz transformation,
the momentum can be tuned to
\mt{(p_{+}, p_{-}, p_{i})=(k_{+}, 0, 0)}
in  light-cone coordinates:
\be
	u^\pm=\tfrac{1}{\sqrt{2}}\,(u^0\pm u^{d-1})\,,
	\qquad i=1,\ldots,d-2\,.
\ee
After solving the same equation for the gauge parameter,
we further gauge fix on-shell to
\be\label{decomp}
\tilde\varphi_{\sst 1,2}^{\sst (s)}(k;u^{+},u^{-},u^{i})
=f^{\sst (s)}_{\sst 1,2}(k;u^{-},u^{i})+
u^+\,(u_i^2)^2\,\tilde{\varphi}_{\sst 1,2}^{\sst (s-5)}(k;u^{+},u^{-},u^{i})\,.
\ee
Note that we cannot gauge fix to zero all the components proportional to $u^{+}$
due to the trace constraints \eqref{tr constr}.
Plugging the solution \eqref{solutionb}
into the gauge condition \eqref{ga con}, yields the two equations:
\ba\label{gc1}
&&k_{+}\,u^+\,\big[\,\pr_{u^-}^2-\tfrac{1}{2}u^+\pr_{u^-}\pr_{u}^2
+\tfrac{1}{8}\,(u^+)^2(\pr_u^2)^2\big]\,\tilde\varphi^{\sst (s)}_{\sst1}+\nonumber\\
&&\quad+\,\big[ \pr_{u^-}-\tfrac{1}{6}\,u^+\pr_u^2+3\,u^+(\pr_{u}^2-\tfrac{1}{2}\,
u^+\pr_{u^-}\pr_{u}^2)+\tfrac{3}{8}\,(u^+)^3(\pr_u^2)^2\big]\,\tilde\varphi^{\sst (s)}_{\sst 2}=0\,,\\
&&\big[\,\pr_{u^-}^2-\tfrac{1}{2}u^+\,\pr_{u^-}\pr_{u}^2+\tfrac{1}{8}(u^+)^3(\pr_u^2)^2\big]\,\tilde\varphi^{\sst (s)}_{\sst 2}=0\,.\label{gc2}
\ea
Using \eqref{decomp} and
$\pr_u^2=\pr_{u^+}\pr_{u^-}+\pr_{u^i}^2$\,, then give
\ba
&\tilde\varphi^{\sst (s-5)}_{\sst 1,2}=0\,,
\qquad
&\quad (\partial_{u^{i}}^{\,2})^{2}f^{\sst (s)}_{\sst 1,2}=0\,,
 \hspace{52pt} \pr_{u^{-}}f^{\sst (s)}_{\sst 2}=0\,,\nn
& \pr_{u^{-}}\,\pr_{u^{i}}^{\,2}\,f^{\sst (s)}_{\sst 1}=0\,,
\qquad
&\pr_{u^{-}}^{\,2} f^{\sst (s)}_{\sst 1}=\tfrac1{12\,k_{+}}\,
\partial_{u^{i}}^{\,2}\,f^{\sst (s)}_{\sst 2}\,,
\qquad\pr_{u^{-}}^{\,3} f^{\sst (s)}_{\sst 1}=0
 \,.
\ea
Finally, one can identify the $so(d-2)$ polarization tensors
of propagating modes:
\ba
	&& f^{\sst (s)}_{\sst 1}(k;u^{-},u^{i})=
	\theta^{\sst (s)}_{\sst 1}(k;u^{i})+
	(u^{i})^{2}\,\theta^{\sst (s-2)}_{\sst 1}(k;u^{i})
	+u^{-}\,\theta^{\sst (s-1)}_{\sst 1}(k;u^{i})+
	(u^{-})^{2}\,\tfrac{d+2s-6}{12\,k_{+}}\,\theta^{\sst (s-2)}_{\sst 2}(k;u^{i})
	\,,\nn
	&& f^{\sst (s)}_{\sst 2}(k;u^{-},u^{i})=
	\theta^{\sst (s)}_{\sst 2}(k;u^{i})+
	(u^{i})^{2}\,\theta^{\sst (s-2)}_{\sst 2}(k;u^{i})\,,
\ea
which correspond to spin $s, s-1, s-2$ massless regular modes and
spin $s, s-2$ massless ghosts.

\bibliographystyle{JHEP}

\begin{thebibliography}{10}

\bibitem{deWit:1979pe}
B.~de~Wit and D.~Z. Freedman, {\it {Systematics of Higher Spin Gauge Fields}},
  {\em Phys.\ Rev.} {\bf D21} (1980) 358.

\bibitem{Fronsdal:1978rb}
C.~Fronsdal, {\it {Massless Fields with Integer Spin}},  {\em Phys.\ Rev.} {\bf
  D18} (1978) 3624.

\bibitem{Francia:2005bu}
D.~Francia and A.~Sagnotti, {\it {Minimal local Lagrangians for higher-spin
  geometry}},  {\em Phys.\ Lett.} {\bf B624} (2005) 93--104,
  [\href{http://xxx.lanl.gov/abs/hep-th/0507144}{{\tt hep-th/0507144}}].

\bibitem{Buchbinder:2007ak}
I.~Buchbinder, A.~Galajinsky, and V.~Krykhtin, {\it {Quartet unconstrained
  formulation for massless higher spin fields}},  {\em Nucl.\ Phys.} {\bf B779}
  (2007) 155--177, [\href{http://xxx.lanl.gov/abs/hep-th/0702161}{{\tt
  hep-th/0702161}}].

\bibitem{Francia:2002aa}
D.~Francia and A.~Sagnotti, {\it {Free geometric equations for higher spins}},
  {\em Phys.\ Lett.} {\bf B543} (2002) 303--310,
  [\href{http://xxx.lanl.gov/abs/hep-th/0207002}{{\tt hep-th/0207002}}].

\bibitem{Bekaert:2005vh}
X.~Bekaert, S.~Cnockaert, C.~Iazeolla, and M.~Vasiliev, {\it {Nonlinear higher
  spin theories in various dimensions}},
  \href{http://xxx.lanl.gov/abs/hep-th/0503128}{{\tt hep-th/0503128}}.

\bibitem{Bekaert:2010hw}
X.~Bekaert, N.~Boulanger, and P.~Sundell, {\it {How higher-spin gravity
  surpasses the spin two barrier: no-go theorems versus yes-go examples}},
  \href{http://xxx.lanl.gov/abs/1007.0435}{{\tt arXiv:1007.0435}}.

\bibitem{Sagnotti:2011qp}
A.~Sagnotti, {\it {Notes on Strings and Higher Spins}},
  \href{http://xxx.lanl.gov/abs/1112.4285}{{\tt arXiv:1112.4285}}.

\bibitem{Stelle:1976gc}
K.~Stelle, {\it {Renormalization of Higher Derivative Quantum Gravity}},  {\em
  Phys.Rev.} {\bf D16} (1977) 953--969.

\bibitem{Bergshoeff:2009hq}
E.~A. Bergshoeff, O.~Hohm, and P.~K. Townsend, {\it {Massive Gravity in Three
  Dimensions}},  {\em Phys.Rev.Lett.} {\bf 102} (2009) 201301,
  [\href{http://xxx.lanl.gov/abs/0901.1766}{{\tt arXiv:0901.1766}}].

\bibitem{Bergshoeff:2009tb}
E.~A. Bergshoeff, O.~Hohm, and P.~K. Townsend, {\it {On Higher Derivatives in
  3D Gravity and Higher Spin Gauge Theories}},  {\em Annals Phys.} {\bf 325}
  (2010) 1118--1134, [\href{http://xxx.lanl.gov/abs/0911.3061}{{\tt
  arXiv:0911.3061}}].

\bibitem{Bergshoeff:2011pm}
E.~A. Bergshoeff, M.~Kovacevic, J.~Rosseel, P.~K. Townsend, and Y.~Yin, {\it {A
  spin-4 analog of 3D massive gravity}},  {\em Class.Quant.Grav.} {\bf 28}
  (2011) 245007, [\href{http://xxx.lanl.gov/abs/1109.0382}{{\tt
  arXiv:1109.0382}}].

\bibitem{Bergshoeff:2012ud}
E.~A. Bergshoeff, J.~Fernandez-Melgarejo, J.~Rosseel, and P.~K. Townsend, {\it
  {On 'New Massive' 4D Gravity}},  {\em JHEP} {\bf 1204} (2012) 070,
  [\href{http://xxx.lanl.gov/abs/1202.1501}{{\tt arXiv:1202.1501}}].

\bibitem{Metsaev:2007fq}
R.~Metsaev, {\it {Ordinary-derivative formulation of conformal low spin
  fields}},  {\em JHEP} {\bf 1201} (2012) 064,
  [\href{http://xxx.lanl.gov/abs/0707.4437}{{\tt arXiv:0707.4437}}].

\bibitem{Maldacena:2011mk}
J.~Maldacena, {\it {Einstein Gravity from Conformal Gravity}},
  \href{http://xxx.lanl.gov/abs/1105.5632}{{\tt arXiv:1105.5632}}.

\bibitem{Lu:2011ks}
H.~Lu, Y.~Pang, and C.~Pope, {\it {Conformal Gravity and Extensions of Critical
  Gravity}},  {\em Phys.Rev.} {\bf D84} (2011) 064001,
  [\href{http://xxx.lanl.gov/abs/1106.4657}{{\tt arXiv:1106.4657}}].

\bibitem{Hyun:2011ej}
S.-J. Hyun, W.-J. Jang, J.-H. Jeong, and S.-H. Yi, {\it {Noncritical
  Einstein-Weyl Gravity and the AdS/CFT Correspondence}},  {\em JHEP} {\bf
  1201} (2012) 054, [\href{http://xxx.lanl.gov/abs/1111.1175}{{\tt
  arXiv:1111.1175}}].

\bibitem{Stelle:1977ry}
K.~Stelle, {\it {Classical Gravity with Higher Derivatives}},  {\em
  Gen.Rel.Grav.} {\bf 9} (1978) 353--371.

\bibitem{Lee:1982cp}
S.~Lee and P.~van Nieuwenhuizen, {\it Counting of states in higher derivative
  field theories},  {\em Phys.Rev.} {\bf D26} (1982) 934.

\bibitem{Buchbinder:1987vp}
I.~Buchbinder and S.~Lyakhovich, {\it {Canonical quantization and local measure
  R$^2$ gravity}},  {\em Class.Quant.Grav.} {\bf 4} (1987) 1487--1501.

\bibitem{Deser:2012qg}
S.~Deser, E.~Joung, and A.~Waldron, {\it {Partial Masslessness and Conformal
  Gravity}},  \href{http://xxx.lanl.gov/abs/1208.1307}{{\tt arXiv:1208.1307}}.

\bibitem{Fradkin:1985am}
E.~S. Fradkin and A.~A. Tseytlin, {\it {Conformal Supergravity}},  {\em Phys.\
  Rept.} {\bf 119} (1985) 233--362.

\bibitem{Metsaev:2007rw}
R.~Metsaev, {\it {Ordinary-derivative formulation of conformal totally
  symmetric arbitrary spin bosonic fields}},  {\em JHEP} {\bf 1206} (2012) 062,
  [\href{http://xxx.lanl.gov/abs/0709.4392}{{\tt arXiv:0709.4392}}].

\bibitem{Marnelius:2008er}
R.~Marnelius, {\it {Lagrangian conformal higher spin theory}},
  \href{http://xxx.lanl.gov/abs/0805.4686}{{\tt arXiv:0805.4686}}.

\bibitem{Shaynkman:2004vu}
O.~Shaynkman, I.~Y. Tipunin, and M.~Vasiliev, {\it {Unfolded form of conformal
  equations in M dimensions and o(M + 2) modules}},  {\em Rev.Math.Phys.} {\bf
  18} (2006) 823--886, [\href{http://xxx.lanl.gov/abs/hep-th/0401086}{{\tt
  hep-th/0401086}}].

\bibitem{Metsaev:2008fs}
R.~Metsaev, {\it {Shadows, currents and AdS}},  {\em Phys.Rev.} {\bf D78}
  (2008) 106010, [\href{http://xxx.lanl.gov/abs/0805.3472}{{\tt
  arXiv:0805.3472}}].

\bibitem{Metsaev:2009ym}
R.~Metsaev, {\it {Gauge invariant two-point vertices of shadow fields, AdS/CFT,
  and conformal fields}},  {\em Phys.\ Rev.} {\bf D81} (2010) 106002,
  [\href{http://xxx.lanl.gov/abs/0907.4678}{{\tt arXiv:0907.4678}}].

\bibitem{Vasiliev:2009ck}
M.~Vasiliev, {\it {Bosonic conformal higher-spin fields of any symmetry}},
  {\em Nucl.Phys.} {\bf B829} (2010) 176--224,
  [\href{http://xxx.lanl.gov/abs/0909.5226}{{\tt arXiv:0909.5226}}].

\bibitem{Bekaert:2012vt}
X.~Bekaert and M.~Grigoriev, {\it {Notes on the ambient approach to boundary
  values of AdS gauge fields}},  \href{http://xxx.lanl.gov/abs/1207.3439}{{\tt
  arXiv:1207.3439}}.

\bibitem{Skvortsov:2007kz}
E.~Skvortsov and M.~Vasiliev, {\it {Transverse Invariant Higher Spin Fields}},
  {\em Phys.Lett.} {\bf B664} (2008) 301--306,
  [\href{http://xxx.lanl.gov/abs/hep-th/0701278}{{\tt hep-th/0701278}}].

\bibitem{Campoleoni:2012th}
A.~Campoleoni and D.~Francia, {\it {Maxwell-like Lagrangians for higher
  spins}},  \href{http://xxx.lanl.gov/abs/1206.5877}{{\tt arXiv:1206.5877}}.

\bibitem{Francia:2012rg}
  D.~Francia,
  {\it {Generalised connections and higher-spin equations}},
  {\it Class.\ Quant.\ Grav.\ }  {\bf 29} (2012) 245003,
  [\href{http://xxx.lanl.gov/abs/1209.4885}{{\tt arXiv:1209.4885}}].


\bibitem{Francia:2007ee}
D.~Francia, {\it {Geometric Lagrangians for massive higher-spin fields}},  {\em
  Nucl.Phys.} {\bf B796} (2008) 77--122,
  [\href{http://xxx.lanl.gov/abs/0710.5378}{{\tt arXiv:0710.5378}}].

\bibitem{Francia:2010qp}
D.~Francia, {\it {String theory triplets and higher-spin curvatures}},  {\em
  Phys.Lett.} {\bf B690} (2010) 90--95,
  [\href{http://xxx.lanl.gov/abs/1001.5003}{{\tt arXiv:1001.5003}}].

\bibitem{Riegert:1984hf}
R.~Riegert, {\it The particle content of linearized conformal gravity},  {\em
  Phys.Lett.} {\bf A105} (1984) 110--112.

\bibitem{Deser:2001us}
S.~Deser and A.~Waldron, {\it {Partial masslessness of higher spins in (A)dS}},
   {\em Nucl.Phys.} {\bf B607} (2001) 577--604,
  [\href{http://xxx.lanl.gov/abs/hep-th/0103198}{{\tt hep-th/0103198}}].

\bibitem{Zinoviev:2001dt}
Y.~Zinoviev, {\it {On massive high spin particles in AdS}},
  \href{http://xxx.lanl.gov/abs/hep-th/0108192}{{\tt hep-th/0108192}}.

\bibitem{Metsaev:2009hp}
R.~Metsaev, {\it {CFT adapted gauge invariant formulation of massive arbitrary
  spin fields in AdS}},  {\em Phys.Lett.} {\bf B682} (2010) 455--461,
  [\href{http://xxx.lanl.gov/abs/0907.2207}{{\tt arXiv:0907.2207}}].

\bibitem{Vasiliev:1990en}
M.~A. Vasiliev, {\it {Consistent equation for interacting gauge fields of all
  spins in (3+1)-dimensions}},  {\em Phys.\ Lett.} {\bf B243} (1990) 378--382.

\bibitem{Vasiliev:2004cp}
M.~Vasiliev, {\it {Higher spin gauge theories in any dimension}},  {\em Comptes
  Rendus Physique} {\bf 5} (2004) 1101--1109,
  [\href{http://xxx.lanl.gov/abs/hep-th/0409260}{{\tt hep-th/0409260}}].

\bibitem{Manvelyan:2010jf}
R.~Manvelyan, K.~Mkrtchyan, W.~Ruhl, and M.~Tovmasyan, {\it {On Nonlinear
  Higher Spin Curvature}},  {\em Phys.Lett.} {\bf B699} (2011) 187--191,
  [\href{http://xxx.lanl.gov/abs/1102.0306}{{\tt arXiv:1102.0306}}].

\bibitem{Fradkin:1989md}
E.~Fradkin and V.~Linetsky, {\it {Cubic interaction in conformal theory of
  integer higher spin fields in four-dimensional space-time}},  {\em Phys.\
  Lett.} {\bf B231} (1989) 97.

\bibitem{Segal:2002gd}
A.~Y. Segal, {\it {Conformal higher spin theory}},  {\em Nucl.\ Phys.} {\bf
  B664} (2003) 59--130, [\href{http://xxx.lanl.gov/abs/hep-th/0207212}{{\tt
  hep-th/0207212}}].

\bibitem{Bekaert:2010ky}
X.~Bekaert, E.~Joung, and J.~Mourad, {\it {Effective action in a higher-spin
  background}},  {\em JHEP} {\bf 1102} (2011) 048,
  [\href{http://xxx.lanl.gov/abs/1012.2103}{{\tt arXiv:1012.2103}}].

\bibitem{Boulanger:2011dd}
N.~Boulanger and P.~Sundell, {\it {An action principle for Vasiliev's
  four-dimensional higher-spin gravity}},
  \href{http://xxx.lanl.gov/abs/1102.2219}{{\tt arXiv:1102.2219}}.

\end{thebibliography}

\providecommand{\href}[2]{#2}\begingroup\raggedright\endgroup

\end{document}